\documentclass[journal]{IEEEtran}

\usepackage{multicol}
\usepackage{color}

\usepackage{ntheorem}
\usepackage{graphicx}      
\usepackage{dsfont}
\usepackage{amssymb,amsmath,amsfonts}
\usepackage{epstopdf}
\usepackage{mathtools}
\usepackage{caption}
\usepackage{setspace}

\makeatletter
\renewcommand*\env@matrix[1][*\c@MaxMatrixCols c]{%
  \hskip -\arraycolsep
  \let\@ifnextchar\new@ifnextchar
  \array{#1}}
\makeatother

\DeclareMathOperator*{\argmin}{arg\,min}
\DeclareMathOperator{\modd}{mod}

\newcommand{\Real}{{\mathds R}} 
\newcommand{\Nat}{{\mathds N}} 

\newcommand{\indep}{\rotatebox[origin=c]{90}{$\models$}}

\newtheorem{definition}{Definition}{}
\newtheorem{corollary}{Corollary}{}
{}
\newtheorem{problem}{Problem}{}
\newtheorem{theorem}{Theorem}{}
\newtheorem{remark}{Remark}{}
\newtheorem{lemma}{Lemma}{}
{}

\begin{document}

\title{On Privacy of Dynamical Systems: An Optimal Probabilistic Mapping Approach\\ (Extended Preprint)}

\author{Carlos~Murguia,
        Iman Shames,
        Farhad Farokhi,
        Dragan Ne\v{s}i\'{c},
        and H. Vincent Poor
\thanks{Carlos Murguia is with the Department
of Mechanical Engineering, Eindhoven University of Technology, Eindhoven, The Netherlands, e-mail: c.g.murguia@tue.nl.}
\thanks{
Iman Shames, Farhad Farokhi, and Dragan Ne\v{s}i\'{c} are with the Department of Electrical and Electronic Engineering, The University of Melbourne, Melbourne, Australia, e-mails: iman.shames@unimelb.edu.au, farhad.farokhi@unimelb.edu.au, and dnesic@unimelb.edu.au.}
\thanks{Vincent Poor is with the Department
of Electrical Engineering, Princeton University, Princeton, NJ 08544 USA, e-mail: poor@princeton.edu.}
\thanks{Manuscript received October 25, 2019; revised October 26, 2019.}}

\markboth{January~2021}%
{On Privacy of Dynamical Systems: An Optimal Probabilistic Mapping Approach}
%



\maketitle

\begin{abstract}
We address the problem of maximizing privacy of stochastic dynamical systems whose state information is released through quantized sensor data. In particular, we consider the setting where information about the system state is obtained using noisy sensor measurements. This data is quantized and transmitted to a (possibly untrustworthy) remote station through a public/unsecured communication network. We aim at keeping (part of) the state of the system private; however, because the network (and/or the remote station) might be unsecure, adversaries might have access to sensor data, which can be used to estimate the system state. To prevent such adversaries from obtaining an accurate state estimate, before transmission, we randomize quantized sensor data using additive random vectors, and send the corrupted data to the remote station instead. We design the joint probability distribution of these additive vectors (over a time window) to minimize the \emph{mutual information} (our privacy metric) between some linear function of the system state (a desired private output) and the randomized sensor data for a \emph{desired level of distortion}--how different quantized sensor measurements and distorted data are allowed to be. We pose the problem of synthesising the joint probability distribution of the additive vectors as a convex program subject to linear constraints. Simulation experiments are presented to illustrate our privacy scheme.
\end{abstract}


\section{Introduction}

In a hyperconnected world, scientific and technological advances have led to an overwhelming amount of user data being collected and processed by hundreds of companies over public networks. Companies mine this data to provide personalized services. However, these new technologies have also led to an alarming widespread loss of privacy in society and vulnerabilities within critical infrastructure -- e.g., power, water, transportation. Depending on adversaries’ resources, opponents may infer critical (private) information about the operation of systems from public data available on the internet and unsecured/public servers and communication networks. A motivating example of this privacy loss is the potential use of data from smart electrical meters by criminals, advertising agencies, and governments, for monitoring the presence and activities of occupants \cite{Poor1,Poor2}. Other examples are privacy loss caused by information sharing in distributed control systems and cloud computing \cite{Huang:2014:CDP:2566468.2566474}; the use of travel data for traffic estimation in intelligent transportation systems \cite{Gruteser}; and data collection and sharing by the Internet-of-Things (IoT) \cite{WEBER201023}, which is, most of the time, done without the user’s informed consent. These privacy concerns show that there is an acute need for privacy preserving mechanisms capable of handling the new privacy challenges induced by an interconnected world, which, in turn, has attracted the attention of researchers from different fields (e.g., computer science, information theory, and control theory) in the broad area of security and privacy of Cyber-Physical Systems (CPSs) -- engineered systems that integrate computation, networking, and dynamic physical components -- see, e.g., \cite{Book_chapt}-\nocite{Farokhi1}\nocite{Farokhi2}\nocite{Jeroem1}\nocite{Cao_Privacy}\nocite{Nima_Privacy}\nocite{FAROKHI3}\nocite{Pappas}\nocite{Wyner}\nocite{Ozarow}\nocite{Fawaz}\nocite{Jerome1}\nocite{Topcu}\nocite{SORIA}\nocite{Geng}\nocite{Dullerud}\nocite{Poor2}\nocite{Takashi_1}\nocite{Takashi_2}\nocite{Takashi_3}\nocite{chaper_privacy_chaos}\nocite{farokhi2019privacy}\cite{Carlos_Iman1}.

{In most engineering applications, information about the state of systems, say $X$, is obtained through sensor measurements. For collection, this information is usually quantized, and then encoded and sent to a remote station for signal processing and decision-making purposes through communication networks. Examples of such systems are numerous: water and electricity consumption meters, traffic monitoring systems, industrial control systems, and so on. If the communication network is public/unsecured and/or the remote station is untrustworthy, adversaries might access and estimate the state of the system. To avoid an accurate state estimation, before transmission, we randomize quantized sensor data using additive random vectors and send the corrupted data to the remote station instead. These vectors are designed to hide (as much as possible) the private part of the state $S$ -- a desired private output modeled as some linear function of the system state, $S=DX$, for some deterministic matrix $D$. Note, however, that it is not desired to overly distort the original sensor data.} We might change the data excessively for practical purposes. Hence, when selecting the additive distorting vectors, we need to take into account the trade-off between \emph{privacy} and \emph{distortion}. As \emph{distortion metric}, we use the \emph{mean squared error} between the original sensor data, $Y$, and its randomized version, $Z = G(Y)$, for some probabilistic mapping $G(\cdot)$. In this manuscript, we follow an information-theoretic approach to privacy. As \emph{privacy metric}, we propose the \emph{mutual information} \cite{Cover}, $I[\tilde{S};Z]$, between a quantized version $\tilde{S}$ of the private output $S$ and the disclosed randomized sensor data $Z = G(Y)$ (over a finite time window). Mutual information $I[V;W]$ between two jointly distributed vectors, $V$ and $W$, is a measure of the statistical dependence between $V$ and $W$ \cite{Cover}. We design the joint probability distribution of the distorting additive vectors to minimize $I[\tilde{S};Z]$ (over a time window), for a \emph{desired level of distortion} -- how different quantized sensor measurements and distorted data are allowed to be. We pose the problem of synthesising the joint probability distribution of these additive vectors as a convex program subject to linear constraints.

Using additive random vectors to increase privacy is common practice. In the context of privacy of databases, a popular approach is differential privacy \cite{Jerome1,Dwork}, where random noise is added to the response of queries so that private information stored in the database cannot be inferred. In differential privacy, because it provides certain privacy guarantees, Laplace noise is usually used \cite{Dwork2}. However, when maximal privacy with minimal distortion is desired, Laplace noise is generally not the optimal solution. This raises the fundamental question: for a given allowable distortion level, what is the noise distribution achieving maximal privacy? This question has many possible answers depending on the particular privacy and distortion metrics being considered and the system configuration \cite{Topcu}-\nocite{SORIA}\nocite{Geng}\cite{Dullerud}. There are also results addressing this question from an information theoretic perspective, where information metrics -- e.g.,  mutual information, entropy, Kullback-Leibler divergence, and Fisher information -- are used to quantify privacy \cite{Poor1,Poor2,Farokhi1,Farokhi2,FAROKHI3,Fawaz,Fawaz2,Lalita}.

In general, if the data to be kept private follows continuous probability distributions, the problem of finding the optimal additive noise to maximize privacy (even without considering distortion) is hard to solve. If a close-form solution for the distribution is desired, the problem amounts to solving a set of nonlinear partial differential equations which, in general, might not have a solution, and even if they do have a solution, it is hard to find \cite{Farokhi1}. This problem has been addressed by imposing some particular structure on the considered distributions or assuming the data to be kept private is deterministic \cite{Farokhi1,SORIA,Geng}.

The authors in \cite{SORIA,Geng} consider deterministic input data sets and treat optimal distributions as distributions that concentrate probability around zero as much as possible while ensuring differential privacy. Under this framework, they obtain a family of piecewise constant density functions that achieve minimal distortion for a given level of privacy. In \cite{Farokhi1}, the authors consider the problem of preserving the privacy of deterministic databases using constrained additive noise. They use the Fisher information and the Cramer-Rao bound to construct a privacy metric between noise-free data and the one with the additive noise and find the probability density function that minimizes it. Moreover, they prove that, in the unconstrained case, the optimal noise distribution minimizing the Fisher information is Gaussian. This observation has also been made in \cite{Cedric} when using mutual information as a measure of privacy.

Most of the aforementioned papers propose optimal continuous distributions assuming deterministic data. However, in a Cyber-Physical-Systems context, the inherent system dynamics and unavoidable system and sensor noise lead to stochastic non-stationary data and thus existing tools do not fit this setting. Here, we identify two possibilities for addressing our problem: 1) we might inject continuous noise to sensor measurements, then quantize the sum, and send it over the unsecured/public network; or 2), the one considered here, quantize sensor measurements, randomize quantized sensor data using additive random vectors with discrete distributions, and send the randomized data over the network. As motivated above, to address the first option, even assuming deterministic sensor data, we have to impose some particular structure on the distributions of the additive noise; and, if sensor data is stochastic, the problem becomes hard to solve (sometimes even untractable). As we prove in this manuscript, if we select the second alternative, under some mild assumptions on the system dynamics and the additive distorting vectors, we can cast the problem of finding the optimal additive vectors as a constrained convex optimization. To the best of the authors knowledge, this problem has not been considered before as it is posed it here.

\begin{figure*}[t]
  \centering
  \includegraphics[scale=.085]{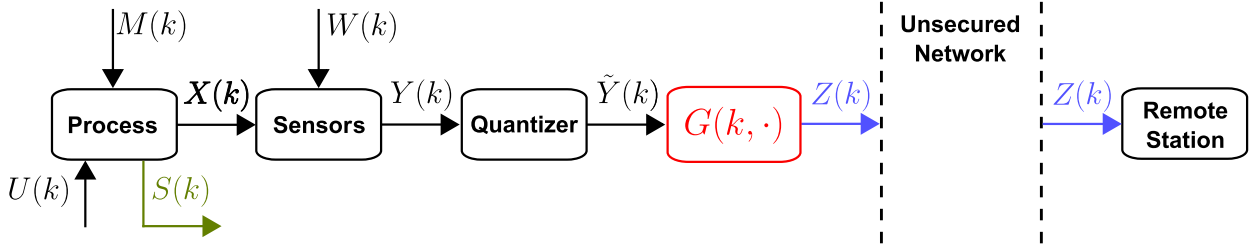}
  \caption{System Configuration.}\label{Fig2}
\end{figure*}

\section{Notation and Preliminaries}\label{Prelim}

\subsection{Notation}
The Euclidian norm in $\Real^n$ is denoted by $||X||$, $||X||^2=X^{\top}X$, where $^{\top}$ denotes transposition. The $n \times n$ identity matrix is denoted by $I_n$ or simply $I$ if $n$ is clear from the context. Similarly, $n \times m$ matrices composed of only ones and only zeros are denoted by $\mathbf{1}_{n \times m}$ and $\mathbf{0}_{n \times m}$, respectively, or simply $\mathbf{1}$ and $\mathbf{0}$ when their dimensions are clear. For positive definite (semidefinite) matrices, we use the notation $P>0$ ($P \geq 0$); moreover, $P>Q$ ($P \geq Q$) means that the matrix $P-Q$ is positive definite (semidefinite). For any two matrices $A$ and $B$, the notation $A \otimes B$ (the Kronecker product \cite{Bollobas}) stands for the matrix composed of submatrices $A_{ij}B$ , where $A_{ij}$, $i,j=1,...,n$, stands for the $ij$th entry of the $n \times n$ matrix $A$. Consider a discrete random vector $X$ with alphabet $\mathcal{X}=\{x_1,\ldots,x_N\}$, $x_i \in \Real^m$, $m,N \in \Nat$, $i\in \{1,\ldots,N\}$, and probability mass function (pmf) $p(x) = \text{Pr}[X=x]$, $x \in \mathcal{X}$, where $\text{Pr}[B]$ denotes probability of event $B$. We denote its probability mass function by $p(x)$ rather than $p_X(x)$ to simplify notation. Thus, $p(x)$ and $p(y)$ refer to two different random vectors, and are, in fact, different probability mass functions, $p_X(x)$ and $p_Y(y)$, respectively. For a discrete stochastic process $X(k)$ taking values from the alphabet $\mathcal{X}_k \subset \Real^m$, we denote its probability mass function as $p_k(x(k))=\text{Pr}[X(k)=x(k)]$, $x(k) \in \mathcal{X}_k$. For simplicity of notation, if the alphabet of $X(k)$ is time-invariant and finite, i.e., $\mathcal{X}_k = \mathcal{X} := \{x_1,\ldots,x_N\}$, $x_i \in \Real^m$, for some finite $m,N \in \Nat$, we write its pmf at time $k$ as $p_k(x)=\text{Pr}[X(k)=x]$, $x \in \mathcal{X}$. We denote by "Simplex" the probability simplex defined by $\sum_{x \in \mathcal{X}}p(x) = 1$, $p(x) \geq 0$ for all $x \in \mathcal{X}$. The notation $X \sim \mathcal{N}[\mu,\Sigma^X]$ means that $X \in \Real^{n}$ is a normally distributed random vector with mean $E[X] = \mu \in \Real^{n}$ and covariance matrix $E[(X-\mu)(X-\mu)^T] = \Sigma^X \in \Real^{n \times n}$, where $E[a]$ denotes the expected value of the random vector $a$. We denote independence between two random vectors, $X$ and $Y$, as $X \indep Y$. Finite sequences of vectors are written as $X^K := (X(1)^{\top},\ldots,X(K)^{\top})^{\top} \in \Real^{Kn}$ and $X^{K_2}_{K_1} := (X(K_1)^{\top},\ldots,X(K_2)^{\top})^{\top} \in \Real^{(K_2-K_1)n}$ with $K_2 > K_1$, $X(i) \in \Real^{n}$, and $n,K,K_1,K_2 \in \Nat$. To avoid confusion, we denote powers of matrices as $(A)^{K} = A \cdots A$ ($K$ times) for $K > 0$, $(A)^{0} = I$, and $(A)^{K} = \mathbf{0}$ for $K < 0$. Given two numbers $a$ and $b$, $b > 0$, the notation $a\modd b$ stands for $a$ modulo $b$, and for a vector $a = (a_1,\ldots,a_n)^\top$, $a_i \in \Real_{>0}$, $i=1,\ldots,n$, $a \modd b = (a_1\modd b,\ldots,a_n\modd b)^\top$.

\subsection{Mutual Information}

\begin{definition}\emph{{\cite{Cover}}}\label{mutual_info}
Consider discrete random vectors, $S$ and $Z$, with joint probability mass function $p(s,z)$ and marginal probability mass functions, $p(s)$ and $p(z)$, respectively. Their mutual information $I[S;Z]$ is defined as the relative entropy between the joint distribution and the product distribution $p(s)p(z)$\emph{:}
$
I[S;Z]:= \sum_{s \in \mathcal{S}}\sum_{z \in \mathcal{Z}}p(s,z)\log \frac{p(s,z)}{p(s)p(z)}.
$
\end{definition}

We use logarithms base 2. Then, mutual information is measured in bits \cite{Cover}.

{\begin{remark}\label{privacy_metric}
Mutual information $I[S;Z]$ between two jointly distributed random vectors, $S$ and $Z$, measures the average amount of information (in bits) about $S$ contained in $Z$ (and vice versa). Then, $I[S;Z]$ can be regarded as a metric of the amount of information about $S$ that is leaked when disclosing $Z$. Mutual information has been widely used as privacy metric, not only for privacy of databases \emph{\cite{Fawaz},\cite{Calmon2},\cite{Sankar}}, but also in a range of applications for dynamical systems \emph{\cite{Poor1},\cite{Poor2},\cite{Takashi_1}-\nocite{Takashi_2}\cite{Takashi_3}}.
\end{remark}}

\section{Problem Formulation}

\subsection{System Description, Quantization, and Stochastic Mappings}

We study discrete-time stochastic systems of the form:
\begin{eqnarray}\label{1}
\left\{ \begin{split}
X(k+1) &= AX(k) + BU(k) +  M(k),\\
Y(k) &= CX(k) + W(k),\\
S(k) &= DX(k),
\end{split} \right.
\end{eqnarray}
with time-instants $k \in \Nat$, state $X \in \Real^{n_x}$, $n_x \in \Nat$, output $Y \in \Real^{n_y}$, $n_y \in \Nat$, \emph{performance (private) output} $S \in \Real^{n_s}$, $n_s \in \Nat$, stochastic disturbances $M \in \Real^{n_x}$ and $W \in \Real^{n_y}$, reference signal $U \in \Real^{n_u}$, $n_u \in \Nat$, and matrices $A \in \Real^{n_x \times n_x}$, $B \in \Real^{n_x \times n_u}$, $C \in \Real^{n_y \times n_x}$, and $D \in \Real^{n_s \times n_x}$. Matrix $D$ is full row rank. The perturbations $M(k)$ and $W(k)$ are i.i.d. multivariate Gaussian processes with $E[M(k)]=\mathbf{0}$, $E[W(k)]=\mathbf{0}$, and covariance matrices $\Sigma^M := E[M(k) M(k)^{\top}] \in \Real^{n_x \times n_x}$, $\Sigma^M > 0$, and $\Sigma^W := E[W(k) W(k)^{\top}] \in \Real^{n_y \times n_y}$, $\Sigma^W > 0$. The initial state $X(1)$ is assumed to be a Gaussian random vector with $E[X(1)]=\mu^X_1 \in \Real^{n_x}$ and covariance matrix $\Sigma^X_1 := E[(X(1)-\mu^X_1)(X(1)-\mu^X_1)^{\top}]  \in \Real^{n_x \times n_x}$, $\Sigma^X_1 > 0$. The processes $M(k)$, $k \in \Nat$ and $W(k)$, $k \in \Nat$, and the initial condition $X(1)$ are mutually independent. It is assumed that the matrices (vectors) $(A,B,C,D,\Sigma^X_1,\mu^X_1,\Sigma^M,\Sigma^W)$ are known, and the reference signal $U(k)$ is known and deterministic.

{\begin{remark}\label{performance_output}
We introduce the notion of private outputs ($S(k)=DX(k)$) for generality. Private outputs  provide freedom to choose the specific part of the state that must be kept private. For instance, if we seek randomizing mechanisms that maximize privacy of the complete state, $D = I_n$; if privacy of the measurable output is required, $D = C$; and if privacy of some other (not necessarily measurable) output, say $R(k) = FX(k)$, is required, $D = F$. For instance, let $X(k) := (P(k),V(k),A(k))^\top \in \mathbb{R}^3$ be a vector consisting on the velocity, position, and acceleration of a vehicle, respectively. Embedded sensors provide position measurements only, i.e., $C = (1,0,0)$. If position is to be kept private, $D = C$, if velocity is the private output, $D = (0,1,0)$, if acceleration must be kept private, $D = (0,0,1)$, and if privacy of the complete state is required, $D = I_3$.
\end{remark}}

Sensor measurements $Y(k) \in \Real^{n_y}$ are quantized using a \emph{vector regular quantizer} \cite{Quantization_Book} $Q_Y(Y(k),N_Y,\mathcal{C},\mathcal{Y})$:
\begin{equation}\label{quantizer}
\tilde{Y}(k) = Q_Y(Y(k),N_Y,\mathcal{C},\mathcal{Y}) :=
\small\left\{
\begin{array}{l}
y_1, $  \hspace{4mm}if $ Y(k) \in c_1, \\ \hspace{20mm} \vdots \\
y_{N_Y}, $ \hspace{1mm}if $ Y(k) \in c_{N_Y},
\end{array}
\right.
\end{equation}
with quantization levels $y_j \in \Real^{n_y}$, $j=1,2,\ldots,N_Y$, quantization cells $c_j \subset \Real^{n_y}$, $\bigcup_j c_j = \Real^{n_y}$, $\bigcap_j c_j = \emptyset$, set of quantization cells $\mathcal{C}:= \{c_1,\ldots,c_{N_Y} \}$, and set of quantization levels $\mathcal{Y}:= \{y_1,\ldots,y_{N_Y} \}$. That is, the vector of quantized sensor measurements, $\tilde{Y}(k) = Q_Y(Y(k),N_Y,\mathcal{C},\mathcal{Y}) \in \mathcal{Y}$, is parametrized by the quantization levels $y_j \in \Real^{n_y}$, the quantization cells $c_j \subset \Real^{n_y}$, $j=1,\ldots,{N_Y}$, and the number of cells ${N_Y} \in \Nat$. Note that, if we know the multivariate probability density $f_k(y(k))$ of $Y(k)$ and the quantizer, we can always obtain the probability mass function $p_k(\tilde{y}(k))$ of $\tilde{Y}(k)$ by integrating $f_k(y(k))$ over the quantization cells $c_j$, $j=1,\ldots,{N_Y}$. Moreover, the alphabet of the \emph{discrete} multivariate random process $\tilde{Y}(k)$ is given by the set of quantization levels $\mathcal{Y}$. Because $\mathcal{Y}$ is time-invariant by construction, we write the pmf of $\tilde{Y}(k)$ as $p_k(\tilde{y})$, $\tilde{y} \in \mathcal{Y}$, i.e., $p_k(\tilde{y}) = \text{Pr}[\tilde{Y}(k)=\tilde{y}]$ for all $\tilde{y} \in \mathcal{Y}$.

\begin{figure}[t]
  \centering
  \includegraphics[scale=.1]{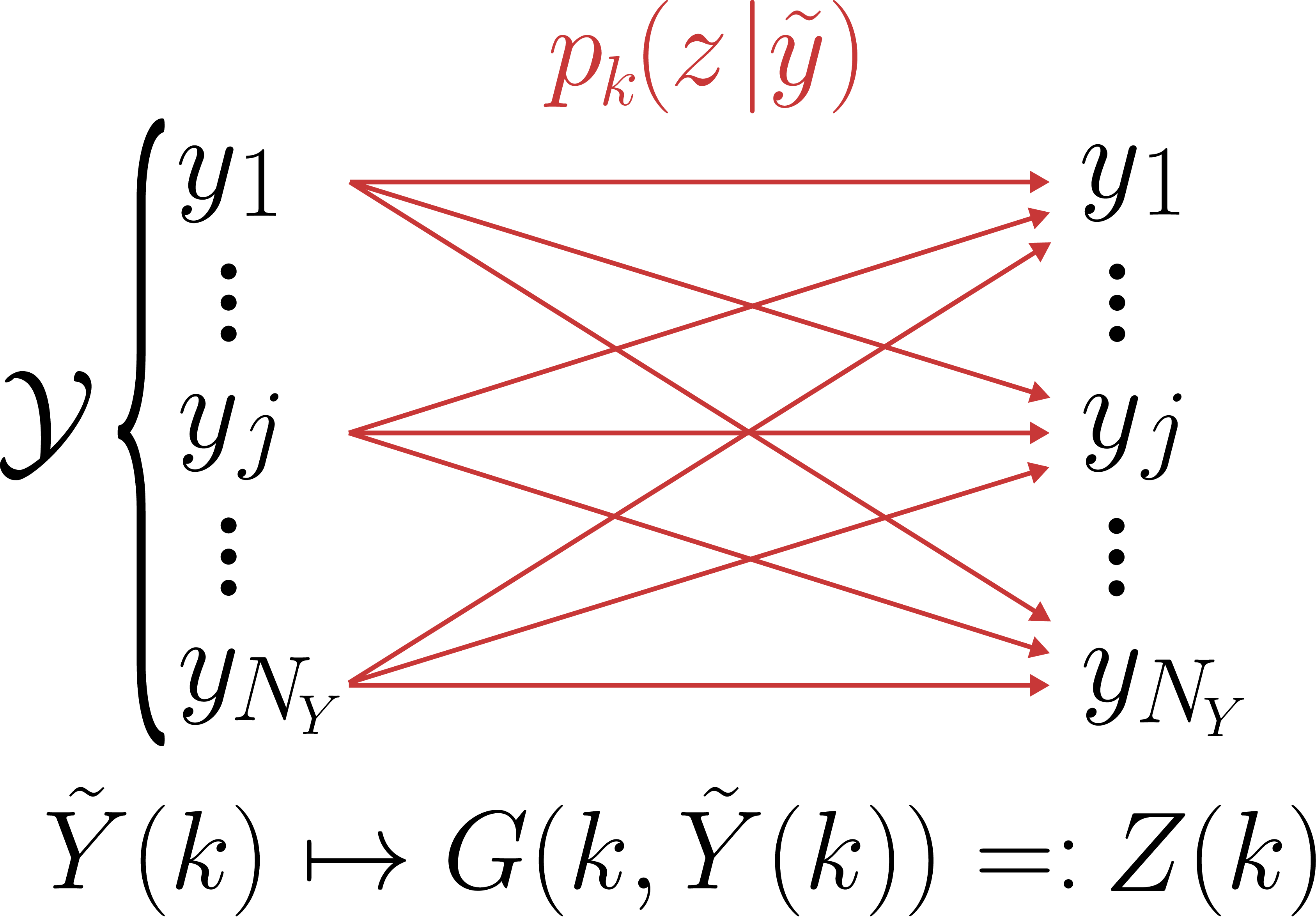}
  \caption{Probabilistic Mapping.}\label{mapping}
\end{figure}

After $Y(k)$ is quantized, we pass $\tilde{Y}(k)$ through a stochastic mapping $G:\Nat \times \mathcal{Y} \rightarrow \mathcal{Y}$ characterized by the transition probabilities $p_k(z|\tilde{y})= \text{Pr}[Z(k) =z|\tilde{Y}(k)=\tilde{y}]$, $\tilde{y},z \in \mathcal{Y}$, i.e., $Z(k) = G(k,\tilde{Y}(k)) \in \mathcal{Y}$, see Figure \ref{mapping}. The vector $Z(k)$ is transmitted over an unsecured communication network to a (possibly untrustworthy) remote station, see Figure \ref{Fig2}. Note that, even by passing $\tilde{Y}(k)$ through $G(k,\cdot)$ before transmission, information about the private output $S(k)$ is directly accessible through $Z(k)$ at the unsecured network. Here, we aim at finding the mapping $G(k,\cdot)$ (the transition probabilities $p_k(z|\tilde{y})$) that minimizes this information leakage. Note, however, that we do not want to make $\tilde{Y}(k)$ and $Z(k)$ overly different either. By passing $\tilde{Y}(k)$ through $G(k,\cdot)$, we might \emph{distort} $\tilde{Y}(k)$ excessively for practical purposes. Hence, when designing the distribution $p_k(z|\tilde{y})$, we need to consider the trade-off between \emph{privacy} and \emph{distortion}.

\begin{remark}\label{remote_station}
  The framework that we propose aims to increase privacy from the system designer point of view without assuming any particular use of the disclosed data $Z(k)$ at the remote station. The vector $Z(k)$ could be used for any real-time application that does not incur in feedback to the system dynamics. In general, applications of this sort are related to remote decision making and monitoring. Particular examples are predictive maintenance, fault-detection, state estimation, and collective decision making.
\end{remark}

{The setting that we consider is a standard privacy-utility tradeoff problem. As we increase privacy (by distorting the original data), the utility of the data (the use of the data at the remote station) would decrease for most applications. The setting that we propose seeks the distorting mechanism that achieve the best privacy-utility tradeoff. This framework makes sense for the potential applications that we introduced in Remark 3, i.e., for remote decision making and monitoring applications. Indeed, if the performance of the application is so critical that no distortion can be tolerated; then, this scheme is not suitable for that particular application. But if there is a small amount of distortion that the application can tolerate (e.g., in fault-detection, a small increase in the false-alarm rate, or in estimation problems, a small increase in the variance of the asymptotic estimation error), then, our scheme looks for the best randomizing mechanism that yields that small amount of allowable distortion while maximizing privacy.}

\vspace{.5mm}

{\begin{remark}\label{untrusty_station}
In the problem setting introduced above, we are assuming that either the communication network or the remote station might be untrustworthy. That is, we aim to prevent adversaries eavesdropping on the channel and those at the remote station from accurately estimating the private output $S(k)$. Standard encryption techniques would not work in this setting as, after decryption, adversaries at the remote station could accurately estimate the private output. There are applications where Partially Homomorphic Encryption (PHE) has been used to do computations over encrypted data and avoid decrypting at the remote station (e.g., control and optimization applications \emph{\cite{FAROKHI201713}\nocite{KIM2016175}\nocite{Gatsis_homo}-\cite{Murguia_Encrypt}}). Note, however, that applications for which PHE can be used are still limited. The issue with using PHE is that the application at the remote station must work on the encrypted data. Meaning that the algorithm that we would normally use without encryption has to be redesigned to work with data in the encrypted domain (usually finite rings of integers). This is quite challenging as it is not always possible to adapt algorithms in such a way. Moreover, techniques based on PHE still suffer from insider attacks, i.e., if someone from the inside distributes the secret key for decryption, adversaries could access undistorted data directly. The scheme that we propose here avoids this by ``destroying'' the private information from the shared distorted data. That is, we inject uncertainty in the direction of the private output $S(k)$ while distorting the original data $\tilde{Y}(k)$ as little as possible. By doing this, we decrease the estimation performance of any adversary (independently of the estimation algorithm they use) as information about $S(k)$ is simply not there, or phrased differently, it is optimally obfuscated.
\end{remark}}

\subsection{Adversarial Model}

{We consider worst-case adversaries that eavesdrop data at the communication network and/or the remote station. They do not only have access to all $Z(k)$, $k \in \mathcal{K}$, but also know the dynamics, quantizer, reference, and stochastically in the system, i.e., matrices  $(A,B,C,D,\Sigma^X_1,\mu^X_1,\Sigma^M,\Sigma^W)$ and the deterministic reference sequence $U(k)$, $k \in \mathcal{K}$, are perfectly known by the adversary.}

{In practice, most of the times, actual adversaries would not have all the capabilities that we assume in this section. However, if we maximize privacy under worst-case adversaries, we ensure that adversaries with less capabilities perform even worse (or equal at most). That is, the privacy guarantees that the scheme provides under worse-case adversaries would hold for weaker ones. These guarantees provide an upper bound on the information that could actually be leaked for any adversary. Assuming worst-case adversaries is common practice in security and privacy of cyber-physical systems, see, e.g., \cite{WEBER201023,Jeroem1,Nima_Privacy,Fawaz,farokhi2019privacy,Fawaz2}, \cite{Poor}\nocite{Pasqualetti2013}\nocite{Mo2016}-\cite{Teixeira2015}. The uncertainty induced by the potential limited knowledge of adversaries is usually pushed aside, and the privacy mechanism focuses on the fundamental privacy leakage from the distorted data generated by the privacy mechanism.}

\begin{figure}[t]
  \centering
  \includegraphics[scale=.0725]{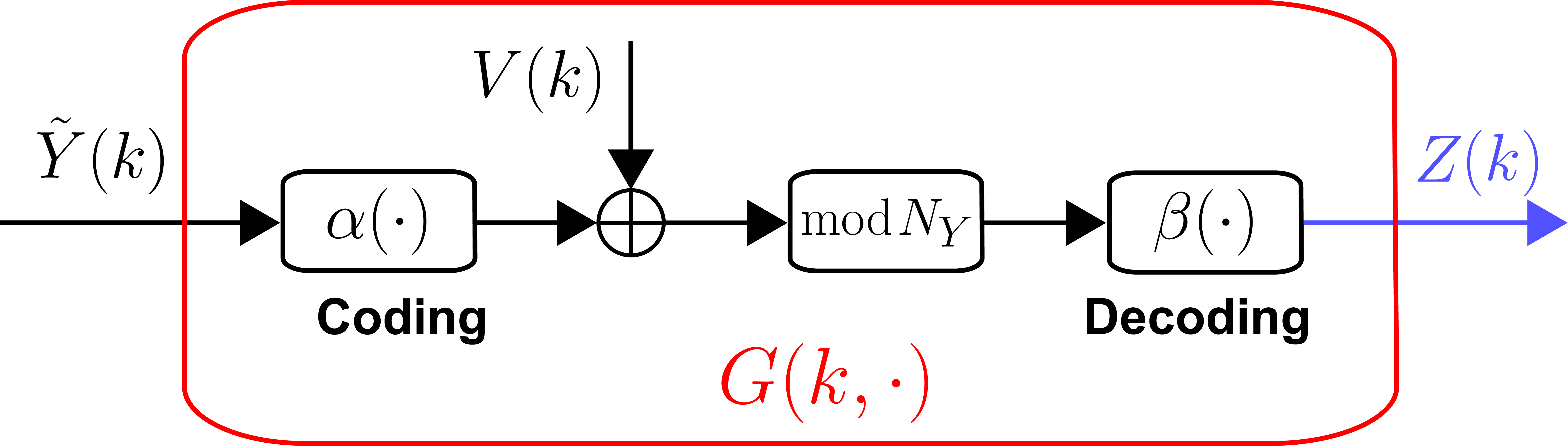}
  \caption{Schematic diagram of the mapping $G(k,\cdot)$.}\label{Map}
\end{figure}

\subsection{Metrics and Problem Formulation} \label{formulation}

For given time horizon $K \in \Nat$, the aim of our privacy scheme is to make inference of the sequence of private vectors $S^K=(S(1)^{\top},\ldots,S(K)^{\top})^{\top}$ from the distorted sequence $Z^K=(Z(1)^{\top},\ldots,Z(K)^{\top})^{\top}$ as hard as possible without distorting $\tilde{Y}^K=(\tilde{Y}(1)^{\top},\ldots,\tilde{Y}(K)^{\top})^{\top}$ excessively. As \emph{distortion metric}, we use the stacked mean squared error: $E[||Z^K - \tilde{Y}^K||^2]$; and, {as \emph{privacy metrics}, the mutual information $I[\tilde{S}^K;Z^K]$, where $\tilde{S}^K = (\tilde{S}(1)^{\top},\ldots,\tilde{S}(K)^{\top})^{\top}$ and $\tilde{S}(i) \in \Real^{n_s}$ denotes a quantized version of the private output $S(i)$, $i=1,\ldots,K$. A natural question arises here: Why don't we directly use $I[S^K;Z^K]$ as privacy metric? Note that, because $X(1)$, $M(k)$, and $W(k)$ are Gaussian, entries of $S^K$ are jointly normally distributed. Moreover, $Z^K$ is a discrete random vector whose conditional distribution depends on the quantizer, $Q_Y(Y(k),N_Y,\mathcal{C},\mathcal{Y})$, and the density of $Y(k)$. Hence, $I[S^K;Z^K]$ would denote the mutual information between continuous and discrete random vectors. This type of mutual information is not always well defined. There are some hybrid definitions available in the literature that have been used to estimate mutual information between continuous and discrete data sets \cite{Torkkola}-\nocite{Kannan}\nocite{RossMutualInf}\nocite{Hanchuan}\cite{HU201110737}. However, we prefer to avoid using hybrid formulations because they lead to complicated expressions that are hard to evaluate. We work directly with the definition of mutual information introduced by Shannon for discrete random vectors (given in Definition \ref{mutual_info}) by discretizing the density of $S(k)$ to compute $I[\tilde{S}^K;Z^K]$. This formulation allows us to obtain suboptimal distorting mechanisms and, as we decrease the size of the discretization cells of $\tilde{S}(k)$ to zero, we recover $I[S^K,Z^K]$ \cite[Chapter 8]{Cover}. We acknowledge that optimality is sacrificed by working with approximations $I[\tilde{S}^K,Z^K]$ of $I[S^K,Z^K]$, but we gain tractability in the sense that we can always cast and solve the synthesis optimization problem when working with $I[\tilde{S}^K,Z^K]$.}

Summarizing the above discussion, we aim at minimizing $I[\tilde{S}^K;Z^K]$ subject to $E[||Z^K - \tilde{Y}^K||^2] \leq \epsilon_K$, for a desired level of distortion $\epsilon_K \in \Real_{>0}$, using as decision variables the conditional probability mass function $p(z^K|\tilde{y}^K) = \text{Pr}[Z^K = z^K|\tilde{Y}^K = \tilde{y}^K]$, where $z^K,\tilde{y}^K \in \mathcal{Y}^K := \mathcal{Y} \times \ldots \times \mathcal{Y}$ ($K$ times) and elements of $\mathcal{Y}^K$ belong to $\Real^{Kn_y}$, i.e., since $\tilde{Y}(k)$ and $Z(k)$ have alphabet $\mathcal{Y}$, $k \in \{1,\ldots,K\}$, the stacked vectors $\mathcal{Y}^K$ and $Z^K$ have alphabet $\mathcal{Y}^K$.

\begin{remark}\label{largeopt1}
{The number of optimization variables $p(z^K|\tilde{y}^K)$ depends on the number of quantization levels $N_Y$ and the horizon $K$. Both stacked vectors $Y^K$ and $Z^K$ take values from the alphabet $\mathcal{Y}^K = \mathcal{Y} \times \ldots \times \mathcal{Y}$ ($K$ times) and because $\mathcal{Y}$ is the set of $N_Y$ quantization levels, $\mathcal{Y}^K$ has $(N_Y)^K$ elements. We aim at computing an optimal transition probability $p(z^K|\tilde{y}^K)$ from each element of the alphabet of $Y^K$ to every element of the alphabet of $Z^K$. Therefore, we have $(N_Y^{K})^2$ optimization variables to minimize $I[\hat{S}^K(\tilde{Y}^K);\hat{S}^K(Z^K)]$ and $I[\tilde{S}^K;Z^K]$. This is already a large-scale optimization problem for quantizers with moderate resolution and fairly small $K$. For instance, a 3-bit quantizer ($N_Y = 8$) and a horizon of $K=5$ requires more than a billion variables. A possible solution to this dimensionality problem is to impose some structure on $p(z^K|\tilde{y}^K)$ to reduce the number of variables. That is, we could set some of the transition probabilities to a constant known value a priori, and use the remaining ones to minimize the mutual information. In what follows, we propose a systematic way to achieve this using an independent additive random process $V(k)$. We impose this structure on the probabilistic mapping $G(k,\cdot)$ (see Figure \ref{Fig2}) so that the number of variables in $p(z^K|\tilde{y}^K)$ is reduced from $(N_Y)^{2K}$ to $(N_V)^{K}$, where $N_V$ denotes the cardinality of the alphabet of additive process $V(k)$ to be designed. Indeed, there is a trade-off between the privacy level that can be achieved (large $N_V$ implies more degrees of freedom to minimize mutual information) and the complexity of the optimization problem to be solved to obtain the privacy mechanism (as increasing the number variables increases the complexity of the problem).}
\end{remark}

The proposed probabilistic mapping $G(k,\cdot)$ consists of the following three objects: 1) a coding function $\alpha: \mathcal{Y} \rightarrow \{0,1,\ldots,N_Y-1\} =:\bar{\mathcal{Y}}$ that indexes each element of $\mathcal{Y}$; 2) a discrete random process $V(k)$, independent of $\tilde{Y}(k)$ for all $k \in \Nat$, with alphabet $\mathcal{V}:= \{0,1,\ldots,N_V-1\}$, $N_V \in \Nat$, and probability mass function $q_k(v)$, $v \in \mathcal{V}$ (the pmf of $V(k)$ is denoted as $q_k(v)$ rather than $p_k(v)$ because, in what follows, we use $q_k(v)$ as optimization variables and we want to clearly distinguish $q_k(v)$ from other probability mass functions); and 3) a decoding function $\beta: \{0,1,\ldots,N_Y-1\} \rightarrow \mathcal{Y}$. We characterize each of these objects before introducing the mapping $G(k,\cdot)$. The indexing (coding) function $\alpha: \mathcal{Y} \rightarrow \bar{\mathcal{Y}}$ is defined as
\begin{equation}\label{Mapping1mod}
\alpha(\zeta) :=
\small\left\{
\begin{array}{l}
0, $  \hspace{11mm}if $ \zeta = y_1, \\ \hspace{21mm} \vdots \\
N_Y-1, $ \hspace{2mm}if $ \zeta = y_{N_Y}.
\end{array}
\right.
\end{equation}
For given $\tilde{Y}(k) \in \mathcal{Y}$ and corresponding $\alpha(\tilde{Y}(k)) \in \{0,1,\ldots,N_Y-1\}$, we add a realization of the process $V(k) \in \{0,1,\ldots,N_V-1\}$ to randomize $\alpha(\tilde{Y}(k))$, and project the sum onto the ring $\{0,1,\ldots,N_Y-1\}$, i.e., $(\alpha(\tilde{Y}(k)) + V_k)\modd N_Y \in \mathcal{Y}$, where $\modd N_Y$ denotes modulo $N_Y$. We project $\alpha(\tilde{Y}(k)) + V_k$ onto $\mathcal{Y}$ to ensure that $Z(k)$ has the same alphabet as $\tilde{Y}(k)$. Then, we decode the sum using the function $\beta: \bar{\mathcal{Y}} \rightarrow \mathcal{Y}$ defined as
\begin{equation}\label{Mapping2mod}
\beta \big(\xi) :=
\small\left\{
\begin{array}{l}
y_1, $  \hspace{5mm}if $ \xi = 0, \\ \hspace{16.25mm} \vdots \\
y_{N_Y}, $ \hspace{2mm}if $ \xi = N_Y-1.
\end{array}
\right.
\end{equation}
Note that $\beta(\alpha(\zeta)) = \zeta$ and $\alpha(\beta(\xi)) = \xi$. We construct the mapping $G:\Nat \times \mathcal{Y} \rightarrow \mathcal{Y}$, $\tilde{Y}(k) \mapsto G(k,\tilde{Y}(k))$, combining  \eqref{Mapping1mod} and \eqref{Mapping2mod} as follows
\begin{align}\label{Mapping3mod}
  Z(k) = G(k,\tilde{Y}(k)) := \beta\big( ( \alpha(\tilde{Y}(k))+V(k) )\modd N_Y \big).
\end{align}
In Figure \ref{Map}, we depict a schematic diagram of the mapping $G(k,\cdot)$. Since $\alpha(\cdot)$ and $\beta(\cdot)$ are fixed injective functions, we can only use the probability mass function $q_k(v) = \text{Pr}[V(k)=v]$, $v \in \mathcal{V} = \{0,1,\ldots,N_V-1\}$, to minimize $I[\hat{S}^K(\tilde{Y}^K);\hat{S}^K(Z^K)]$ and $I[\tilde{S}^K;Z^K]$. For given time horizon $K \in \Nat$, let $V^K=(V(1)^{\top},\ldots,V(K)^{\top})^{\top}$, and denote its probability mass function as $q(v^K) = \text{Pr}[V^K = v^K]$, where $v^K \in \mathcal{V}^K := \mathcal{V} \times \ldots \times \mathcal{V}$ ($K$ times) and elements of $\mathcal{V}^K$ belong to $\Real^{K}$, i.e., since $V(k)$ has alphabet $\mathcal{V}$, $k \in \{1,\ldots,K\}$, the stacked vector $V^K$ has alphabet $\mathcal{V}^K$. In what follows, we formally present the optimization problems we seek to address.

\begin{problem}
Given the system dynamics \eqref{1}, sensor quantizer \eqref{quantizer}, time horizon $K \in \Nat$, desired distortion level $\epsilon_K \in \Real_{\geq 0}$, quantized version $\tilde{S}(k)$ of $S(k)$, $k \in \{1,\ldots,K\}$, and the probabilistic mapping \eqref{Mapping1mod}-\eqref{Mapping3mod}, find the probability mass function $q(v^K)$ solution of the following optimization problem:
\begin{equation} \label{eq:convex_optimization2}
\left\{\begin{aligned}
	&\min_{q(v^K)}\
    I[\tilde{S}^K;Z^K],\\
    &\hspace{4mm}\text{\emph{s.t. }} E[||Z^K - \tilde{Y}^K||^2] \leq \epsilon_K,\\[1mm]
    &\hspace{10mm} V^K \indep \hspace{.5mm} \tilde{Y}^K, \text{ \emph{and} } q(v^K) \in \text{ \emph{Simplex}}.
\end{aligned}\right.
\end{equation}
\end{problem}

\subsection{On the Distortion Constraint and Linear Estimators}

Problem 1 is posed in terms of a general distortion constraint that upper bounds the second moment of $(Z^K - \tilde{Y}^K)$. As we have motivated before (see Remark 3 and the discussion below it), this general formulation seeks to avoid assuming a particular application of the transmitted data at the remote station. If a particular application was considered (e.g., state estimation, fault detection, etc.), we would have to cast the problem in terms of the performance degradation \emph{on that particular application}. Then, for every application, we would have a different performance criteria  -- leading to an endless number of different formulations.\\
Arguably, for the class of linear systems considered in this manuscript, and most applications at the remote station (for this class), performance degradation induced by the privacy mechanism can be written (or upper bounded) in terms of the distortion on $(Z^K - \tilde{Y}^K)$. To illustrate the later, we provide a brief application example using a class of linear state estimators. Consider system \eqref{1} and the following two recursive estimators, one driven by the original quantized sensors, $\tilde{Y}(k)$, and the other driven by the distorted $Z(k)$:
\begin{equation}\label{estm1}
\hat{X}_{Y}(k+1) = A\hat{X}_{Y}(k) + BU(k) + L(\tilde{Y}(k) - C\hat{X}_{Y}(k)),
\end{equation}
\begin{equation}\label{estm2}
\hat{X}_{Z}(k+1) = A\hat{X}_{Z}(k) + BU(k) + L(Z(k) - C\hat{X}_{Z}(k)),
\end{equation}
with corresponding estimator state $\hat{X}_{Y}(k),\hat{X}_{Z}(k) \in \Real^{n_x}$, initial condition $\hat{X}_{Y}(1) = \hat{X}_{Z}(1) = E[X(1)] = \mu^X_1$, reference signal $U(k) \in \Real^{n_u}$ and matrices $(A,B,C)$ as in \eqref{1}, and gain matrix $L \in \Real^{n_x \times n_y}$. Matrix $L$ is designed to achieve the desired estimation performance (e.g., internal stability, disturbances attenuation, and convergence rate). Estimators \eqref{estm1} and \eqref{estm2} are identical in structure but driven by different sequences. This class of estimators is referred in the literature as Luenberger observers \cite{Astrom}. Here, we present a prescriptive analysis where we compare the performance between the estimators, for $k \in \mathcal{K}$, and show that the performance degradation induced by the use of the distorted $Z(k)$, instead of the original $\tilde{Y}(k)$, is upper bounded by a convex function of $E[||Z^K - \tilde{Y}^K||^2]$.\\
Define the estimation errors $e^Y(k) := X(k) - \hat{X}^Y(k)$ and $e^Z(k) := X(k) - \hat{X}^Z(k)$. Given the system and estimators dynamics, \eqref{1} and \eqref{estm1}-\eqref{estm2}, it is easy to verify that the estimation errors evolve as
\begin{align*}
&e_{Y}(k+1) = \bar{A}e_{Y}(k) + M(k) - L(W(k) + \delta^Y(k)),\\
&e_{Z}(k+1) = \bar{A}e_{Y}(k) + M(k) - L(W(k) + \delta^Y(k) - \delta^Z(k)),
\end{align*}
with $\bar{A} := A-LC$, quantization error $\delta^Y(k):= Y(k) - \tilde{Y}(k)$,\linebreak and $\delta^Z(k):= Z(k) - \tilde{Y}(k)$. Matrix $L$ is selected such that $\rho(A-LC) < 1$, where $\rho(\cdot)$ denotes spectral radius \cite{Horn}. Such a matrix $L$ exists when system \eqref{1} is \emph{detectable} \cite{Astrom}. The condition $\rho(A-LC) < 1$ guarantees that, if disturbances $(M(k),W(k),\delta^Y(k),\delta^Z(k))$ are equal to zero (or converge to zero asymptotically) $\lim_{k \rightarrow \infty} e^Y(k) = \lim_{k \rightarrow \infty} e^Z(k) = 0$ (internal stability) \cite{Astrom}. For non-vanishing disturbances, the estimation errors do not converge to zero. They converge to a compact invariant set, if disturbances are uniformly bounded on compact sets, and to a stationary process, if disturbances are stationary random processes \cite{Astrom}.\\
The performance of \eqref{estm1} and \eqref{estm2} is fully characterized by the estimation errors $e^Y(k)$ and $e^Z(k)$. Thus, to compare their performance, we introduce the auxiliary vector $\Delta(k) := E[e^Y(k) - e^Z(k)]$. Given the difference equations for $e^Y(k)$ and $e^Z(k)$ introduced above, it is easy to verify that $\Delta(k)$ evolves as $\Delta(k+1) = \bar{A}\Delta(k) + LE[\delta^Z(k)]$, with $\Delta(1) = \mathbf{0}$. Hence, the general solution of $\Delta(k)$ is given by
\begin{equation}\label{estm3}
\Delta(k) = \sum_{i=0}^{k-2}\bar{A}^iLE[\delta^Z(k-i-1)], \hspace{1mm}k>1.
\end{equation}
Note that $\delta^Z(k) = \mathbf{0}$, $k \in \Nat$, implies $\Delta(k) = \mathbf{0}$, $k \in \Nat$, because $\Delta(1) = \mathbf{0}$. That is, the auxiliary vector is nonzero only if there is distortion due to the privacy mechanism. The aim of this subsection is to show that an upper bound $\epsilon_K$ on $E[||Z^K - \tilde{Y}^K||^2]$ translates into an upper bound on the estimation performance degradation (i.e., into an upper bound on $||\Delta(k)||$). To accomplish this, we seek to upper bound $||\Delta^{K+1}||$ by a convex function of $E[||Z^K - \tilde{Y}^K||^2]$.\\
For $k \in \mathcal{K}$, equation \eqref{estm3} can be written in terms of $Z^K$, $\tilde{Y}^K$, and $\Delta^{K+1} = (\Delta(1)^{\top},\ldots,\Delta(K+1)^{\top})^{\top}$ as
\begin{equation}\label{estm4}
\Delta(K+1) = \Theta_K(I_{K-1} \otimes L)E[Z^K - \tilde{Y}^K],
\end{equation}
with
\begin{equation*}
  \Theta_K = \begin{bmatrix} \mathbf{0} & \mathbf{0} & \mathbf{0} & \cdots & \mathbf{0} \\ I & \mathbf{0} & \mathbf{0} & \cdots & \mathbf{0} \\ \bar{A} & I & \mathbf{0} & \cdots & \mathbf{0} \\ \vdots & \vdots & \vdots & \ddots & \vdots \\[1mm] (\bar{A})^{K-2} & (\bar{A})^{K-3} & (\bar{A})^{K-4} & \cdots & I  \end{bmatrix}.
\end{equation*}
It follows that
\begin{align*}
||\Delta(K+1)|| &\overset{\text{(a)})}{\leq}  \vartheta || E[Z^K - \tilde{Y}^K] ||,\\
                &\overset{\text{(b)})}{\leq} \vartheta  E[||Z^K - \tilde{Y}^K||],\\
                &\overset{\text{(c)})}{\leq} \vartheta  E[||Z^K - \tilde{Y}^K||^2]^{1/2},\\
                &\overset{\text{(d)})}{\leq} \vartheta  \sqrt{\epsilon_K},
\end{align*}
where $|| \cdot ||$ denotes Euclidian norm for vectors and spectral  norm for matrices \cite{Horn}, $\vartheta := ||\Theta_K(I_{K-1} \otimes L)||$, (a) follows from properties of matrix norms, (b) and (c) from Jensen's inequality \cite{Ross}, and (d) from \eqref{eq:convex_optimization2}. Therefore, by solving \eqref{eq:convex_optimization2} with the distortion constraint, $E[||Z^K - \tilde{Y}^K||^2] \leq \epsilon_K$, we enforce that the estimation performance is degraded by at most $\vartheta  \sqrt{\epsilon_K}$, with $\vartheta$ being a positive constant that is independent of the privacy mechanism. Similar prescriptive analyses can be performed for other applications at the remote station (e.g., fault detection and isolation, distributed decision-making and optimization, predictive maintenance, classification, filtering, etc). By casting \eqref{eq:convex_optimization2} in terms of $E[||Z^K - \tilde{Y}^K||^2] \leq \epsilon_K$, we cover a variety of applications as their performance degradation can be written (or upper bounded) in terms of $E[||Z^K - \tilde{Y}^K||^2]$. Hence, hereafter, we focus on Problem 1 as it is casted in \eqref{eq:convex_optimization2}.

\section{Solution to Problem 1} \label{finite_horizon}

Consider the cost function $I[\tilde{S}^K;Z^K]$ for some quantized version $\tilde{S}^K$ of the stacked private output $S^K$, and the system dynamics \eqref{1}. At time $k$, $S(k)$ can be written in terms of the initial condition, the stacked reference $U^{k-1}$, and the stacked disturbance $M^{k-1}$ as $S(k) = DA^{k-1}X(1) + \sum_{i=0}^{k-2}DM(k-1-i)+DU(k-1-i)$, i.e., $S(k)$ is the sum of independent Gaussian random vectors (and thus it is also Gaussian). It follows that the support of $S(k)$ is the whole $R^{n_s}$. To discretize the density of $S(k)$, we divide its support into a finite set of cells. Let $\mathcal{H}:= \{h_1,\ldots,h_{N_S} \}$ denote a set of $N_S \in \Nat$ quantization cells satisfying: $h_j \subset \Real^{n_s}$, $\bigcup_j h_j = \Real^{n_s}$, $\bigcap_j h_j = \emptyset$. Associated with each cell $h_j$, we introduce the corresponding quantization level $s_j \in h_j \subset \Real^{n_s}$, i.e., $s_j$ denotes a $n_s$-dimensional point in the interior of $h_j$, $j \in \{1,\ldots,N_S\}$. We collect all the quantization levels into the set $\mathcal{S} := \{s_1,\ldots,s_{N_S}\}$. Then, the quantized  private output, $\tilde{S}(k) \in \Real^{n_s}$, can be written as
\begin{equation}\label{quantizerX}
\tilde{S}(k) = Q_S(S(k),N_S,\mathcal{H},\mathcal{S}) :=
\small\left\{
\begin{array}{l}
s_1, $  \hspace{2.5mm}if $ S(k) \in h_1, \\ \hspace{18mm} \vdots \\
s_{N_S}, $ if $ S(k) \in h_{N_S}.
\end{array}
\right.
\end{equation}
Denote the stacked vector of quantized private outputs as $\tilde{S}^K = (\tilde{S}(1)^\top,\ldots,\tilde{S}(K)^\top)^\top \in R^{Kn_s}$. Note that if we know the joint probability density function of $S^K$, once we have fixed the quantizer $Q_S(\cdot)$, we can obtain the probability mass function $p(\tilde{s}^K) = \text{Pr}[\tilde{S}^K = \tilde{s}^K]$, $\tilde{s}^K \in \mathcal{S}^K := \mathcal{S}\times \ldots \times \mathcal{S}$ ($K$ times) by integrating the density of $S^K$ over the cells in $\mathcal{H}$. For the discrete random vector $\tilde{S}^K$, we have a well defined mutual information $I[\tilde{S}^K;Z^K]$. By Definition \ref{mutual_info}, the cost $I[\tilde{S}^K;Z^K]$ is a function of $p(\tilde{s}^K,z^K)$, and the marginals $p(\tilde{s}^K)$ and $p(z^K)$. However, to minimize $I[\tilde{S}^K;Z^K]$, we need to write it in terms of $q(v^K)$ (our design variables). Notice that, if the joint density of $S^Y$ and $Y^K$ is a non-degenerate Gaussian, we can numerically compute $p(\tilde{s}^K|\tilde{y}^K)$ for any $\tilde{s}^K \in \mathcal{S}^K$ and $\tilde{y}^K \in \mathcal{Y}^K$; and that, because (by construction) $\tilde{S}(k)$ and $Z(k)$ are conditionally independent given $\tilde{Y}(k)$, $\tilde{S}^K$ and $Z^K$ are conditionally independent given $\tilde{Y}^K$. The latter implies that $p(\tilde{s}^K,z^K) = \sum_{\tilde{y}^K \in \mathcal{Y}^K}p(\tilde{y}^K)p(\tilde{s}^K|\tilde{y}^K)p(z^K|\tilde{y}^K)$. Then, for given $p(\tilde{y}^K)$ and $p(\tilde{s}^K|\tilde{y}^K)$, we can write the cost $I[\tilde{S}^K;Z^K]$ in terms of $p(z^K|\tilde{y}^K)$ and $q(v^K)$. In the following lemma, we write the cost function $I[\tilde{Y}^K;Z^K]$ in terms of $q(v^K)$, and prove that it is convex in $q(v^K)$. Before stating the lemma, we need some extra notation. For any $\zeta^K = (\zeta(1)^\top,\ldots,\zeta(K)^\top)^\top \in \mathcal{Y}^K$, $\zeta(i) \in \mathcal{Y}$, $i \in \{1,\ldots,K\}$, define the stacked indexing function $\bar{\alpha}:\mathcal{Y}^K \rightarrow \bar{\mathcal{Y}}^K$, with $\bar{\mathcal{Y}}^K := \bar{\mathcal{Y}} \times \cdots \times \bar{\mathcal{Y}}$ ($K$ times) and $\bar{\mathcal{Y}} = \{0,\ldots,N_Y-1\}$, as:\\
\begin{equation}\label{stacked_alpha}
\bar{\alpha}(\zeta^K) := (\alpha(\zeta(1)),\ldots,\alpha(\zeta(K)))^\top,
\end{equation}
where $\alpha(\cdot)$ is the indexing function defined in \eqref{Mapping1mod}.

\begin{lemma}\label{lem6}
$I[\tilde{S}^K;Z^K]$ is a convex function of $q(v^K)$ for given $p(\tilde{s}^K|\tilde{y}^K)$ and $p(\tilde{y}^K)$, and is written as:
\begin{subequations}
\begin{align}
&I[\tilde{S}^K;Z^K] = \sum_{z^K \in \mathcal{Y}^K}\sum_{\tilde{s}^K \in \mathcal{S}^K}p(\tilde{s}^K)p(z^K|\tilde{s}^K)\log \frac{p(z^K|\tilde{s}^K)}{p(z^K)},\label{cost3}\\[1mm]
&p(z^K|\tilde{s}^K) = \sum_{\tilde{y}^K \in \mathcal{Y}^K}p(\tilde{y}^K|\tilde{s}^K)p(z^K|\tilde{y}^K),\label{cost3a}\\[1mm]
&p(z^K) = \sum_{\tilde{s}^K \in \mathcal{S}^K} \sum_{\tilde{y}^K \in \mathcal{Y}^K}p(\tilde{s}^K)p(\tilde{y}^K|\tilde{s}^K)p(z^K|\tilde{y}^K),\label{cost3b}\\[1mm]
&p(z^K|\tilde{y}^K) = q\big( (\bar{\alpha}(z^K) - \bar{\alpha}(\tilde{y}^K))\modd N_Y \big),\label{cost3c}
\end{align}
\end{subequations}
where $\bar{\alpha}:\mathcal{Y}^K \rightarrow \bar{\mathcal{Y}}^K$ denotes the stacked indexing function defined in \eqref{stacked_alpha}.\\[1mm]
\emph{\textbf{\emph{Proof}}: See Appendix A.}
\end{lemma}

By Lemma \ref{lem6}, $I[\tilde{S}^K;Z^K]$ is convex in our decision variables $q(v^K)$, for given $p(\tilde{s}^K)$ and $p(\tilde{y}^K|\tilde{s}^K)$. Then, if the distortion constraint, $E[||Z^K - \tilde{Y}^K||^2] \leq \epsilon_K$, is convex in $q(v^K)$, and we know $p(\tilde{s}^K)$ and $p(\tilde{y}^K|\tilde{s}^K)$, we could minimize $I[\tilde{S}^K;Z^K]$ efficiently using off-the-shelf optimization algorithms.

\begin{lemma}\label{lem3}
$E[||Z^K - \tilde{Y}^K||^2]$ is a linear function of $q(v^K)$ for given $p(\tilde{y}^K)$, and can be written as follows:
\begingroup\makeatletter\def\f@size{9.5}\check@mathfonts
\def\maketag@@@#1{\hbox{\m@th\normalsize\normalfont#1}}%
\begin{subequations}\label{distortion}
\begin{align}
&E[||Z^K - \tilde{Y}^K||^2] = \sum_{\tilde{y}^K \in \mathcal{Y}^K}\sum_{z^K \in \mathcal{Y}^K} p(z^K,\tilde{y}^K)\big(z^K - \tilde{y}^K \big)^2,\\
&p(z^K,\tilde{y}^K) = q\big( (\bar{\alpha}(z^K) - \bar{\alpha}(\tilde{y}^K))\modd N_Y \big) p(\tilde{y}^K).
\end{align}
\end{subequations}\endgroup
\emph{\textbf{\emph{Proof}}: See Appendix B.
}\end{lemma}

By Lemma \ref{lem6} and Lemma \ref{lem3}, the cost, $I[\tilde{S}^K;Z^K]$, and constraint, $E[||Z^K - \tilde{Y}^K||^2] \leq \epsilon_K$, are parametrized by $p(\tilde{s}^K)$ and $p(\tilde{y}^K|\tilde{s}^K)$. To obtain these distributions, we need to integrate the density of $S^K$ and the joint density of $(S^K,Y^K)$ (if they are not degenerate) over the quantization cells. By lifting the system dynamics \eqref{1} over $\{1,\ldots,K\}$, we can write the stacked vector $((Y^K)^\top,(S^K)^\top)^\top \in \Real^{K(n_s + n_y)}$ as
\begin{align}\label{stackedXY}
&\begin{bmatrix} Y^K \\ S^K \end{bmatrix} = \begin{bmatrix} \tilde{C}_K  \\ \tilde{D}_K \end{bmatrix}F_K X(1) + \begin{bmatrix} \tilde{C}_K  \\ \tilde{D}_K \end{bmatrix} T_K M^{K-1}\\
&\hspace{15mm} + \begin{bmatrix} \tilde{C}_K  \\ \tilde{D}_K \end{bmatrix} L_K U^{K-1} + \begin{bmatrix} I \\ \mathbf{0} \end{bmatrix}W^{K}, \notag
\end{align}
with $L_K := T_K(I_{K-1} \otimes B)$, $\tilde{C}_K := I_K \otimes C$, $\tilde{D}_K := I_K \otimes D$, and
\begin{equation}\label{stackedY}
\left\{ \begin{aligned}
  F_K &:= \begin{bmatrix} I & A^\top & \hdots & (A^\top)^{K-1}  \end{bmatrix}^\top,\\
  T_K &:= \begin{bmatrix} \mathbf{0} & \mathbf{0} & \mathbf{0} & \cdots & \mathbf{0} \\ I & \mathbf{0} & \mathbf{0} & \cdots & \mathbf{0} \\ A & I & \mathbf{0} & \cdots & \mathbf{0} \\ \vdots & \vdots & \vdots & \ddots & \vdots \\[1mm] (A)^{K-2} & (A)^{K-3} & (A)^{K-4} & \cdots & I  \end{bmatrix},
\end{aligned} \right.
\end{equation}
\begin{lemma}\label{stackedDist}
\[ \begin{psmallmatrix} Y^K \\ S^K  \end{psmallmatrix} \sim \mathcal{N}\left[\begin{bsmallmatrix} \tilde{C}_K  \\ \tilde{D}_K \end{bsmallmatrix}F_K \mu^X_1 + \begin{psmallmatrix} \tilde{C}_K  \\ \tilde{D}_K \end{psmallmatrix} L_K U^{K-1},\Sigma^{Y,S}_K\right],\] with covariance $\Sigma^{Y,S}_K \in \Real^{K(n_s + n_y) \times K(n_s + n_y)}$, $\Sigma^{Y,S}_K>0$\emph{:}
\begin{equation}\label{stackeconmatrix3}
\begingroup
\renewcommand*{\arraycolsep}{2pt}
\begin{aligned}
  \Sigma^{Y,S}_K := &\begin{bsmallmatrix} I \\ \mathbf{0} \end{bsmallmatrix}(I_{K} \otimes \Sigma^W)\begin{bsmallmatrix} I \\ \mathbf{0} \end{bsmallmatrix}^\top + \begin{bsmallmatrix} \tilde{C}_K  \\ \tilde{D}_K \end{bsmallmatrix} F_K \Sigma^X_1 F_K^\top \begin{bsmallmatrix} \tilde{C}_K  \\ \tilde{D}_K \end{bsmallmatrix}^\top\\ &+ \begin{bsmallmatrix} \tilde{C}_K  \\ \tilde{D}_K \end{bsmallmatrix}T_K (I_{K-1} \otimes \Sigma^M)T_K^\top \begin{bsmallmatrix} \tilde{C}_K  \\ \tilde{D}_K \end{bsmallmatrix}^\top.
\end{aligned}
\endgroup
\end{equation}
\begin{table*}[!ht]
\noindent\rule{\hsize}{1pt}
\begin{equation} \label{eq:convex_optimization15}
\left\{\begin{aligned}
	&\min_{p(z^K|\tilde{y}^K)}\                                                                  \sum_{\tilde{s}^K \in \mathcal{S}^K}\sum_{z^K \in \mathcal{Y}^K}p(\tilde{s}^K) \sum_{\tilde{y}^K \in \mathcal{Y}^K}p(\tilde{y}^K|\tilde{s}^K)p(z^K|\tilde{y}^K)\log \dfrac{\sum_{\tilde{y}^K \in \mathcal{Y}^K}p(\tilde{y}^K|\tilde{s}^K)p(z^K|\tilde{y}^K)}{\sum_{\tilde{s}^K \in \mathcal{S}^K} \sum_{\tilde{y}^K \in \mathcal{Y}^K}p(\tilde{s}^K)p(\tilde{y}^K|\tilde{s}^K)p(z^K|\tilde{y}^K)},\\
    &\hspace{4mm}\text{\emph{s.t. }} \sum_{\tilde{y}^K \in \mathcal{Y}^K}\sum_{z^K \in \mathcal{Y}^K}  \sum_{\tilde{s}^K \in \mathcal{S}^K}p(\tilde{s}^K)p(\tilde{y}^K|\tilde{s}^K)p(z^K|\tilde{y}^K) \left(z^K - \tilde{y}^K \right)^2 \leq \epsilon_K,\\[1mm]
    &\hspace{12mm}p(z^K|\tilde{y}^K) = q\big( (\bar{\alpha}(z^K) - \bar{\alpha}(\tilde{y}^K))\modd N_Y \big),\\[1mm]
    &\hspace{12mm}\text{and, } q(v^K) \in \text{ \emph{Simplex}}.
\end{aligned}\right.
\end{equation}
\noindent\rule{\hsize}{1pt}
\end{table*}
\emph{\textbf{\emph{Proof}}: See Appendix C.
}\end{lemma}

To obtain the density of $S^K$, we marginalize the joint density $\mathcal{N}[\begin{bsmallmatrix} \tilde{C}_K  \\ \tilde{D}_K \end{bsmallmatrix}F_K \mu^X_1 + \begin{psmallmatrix} \tilde{C}_K  \\ \tilde{D}_K \end{psmallmatrix} L_K U^{K-1},\Sigma^{Y,S}_K]$ over $Y^K$. We give this density in the following corollary of Lemma \ref{stackedDist}.

\begin{corollary}
$S^{K} \sim \mathcal{N}[\tilde{D}_KF_K \mu^X_1 + \tilde{D}_KL_KU^{K-1},\Sigma^S_K]$ with covariance $\Sigma^S_K = \begin{psmallmatrix} \mathbf{0}  \\ I_{n_s} \end{psmallmatrix}^\top \Sigma^{Y,S}_K \begin{psmallmatrix} \mathbf{0} \\ I_{n_s}  \end{psmallmatrix} \in \Real^{Kn_s \times Kn_s}$.
\end{corollary}

\begin{remark}\label{jointXY}
Having the joint density of $Y^K$ and $S^K$ allows us to compute, for any set of quantization cells $\{ c_l,\ldots,c_i \}$, $l,\ldots,i \in \{1,\ldots,N_Y\}$ and $\{ h_r,\ldots,h_j \}$, $r,\ldots,j \in \{1,\ldots,N_S\}$, $\text{\emph{Pr}}[ S(1) \in h_r,\ldots,S(K) \in h_j]$ and $\text{\emph{Pr}}[Y(1) \in c_l,\ldots,Y(K) \in c_i | S(1) \in h_r,\ldots,S(K) \in h_j]$. By definition, for $\tilde{s}(1),\ldots,\tilde{s}(K) \in \mathcal{S}$, and $\tilde{y}(1),\ldots,\tilde{y}(K) \in \mathcal{Y}$, $p(\tilde{y}^K|\tilde{s}^K) = \text{\emph{Pr}}[\tilde{Y}(1) = \tilde{y}(1),\ldots,\tilde{Y}(K) = \tilde{y}(K) | \tilde{S}(1) = \tilde{s}(1),\ldots,\tilde{S}(K) = \tilde{s}(K)]$. Moreover, by construction of the quantizers, we have:
\begingroup\makeatletter\def\f@size{9.5}\check@mathfonts
\def\maketag@@@#1{\hbox{\m@th\normalsize\normalfont#1}}%
\begin{align*}
&\text{\emph{Pr}}[\tilde{Y}(1) = y_l,\ldots,\tilde{Y}(K) = y_i | \tilde{S}(1) = s_r,\ldots,\tilde{S}(K) = s_j]\\
&= \text{\emph{Pr}}[Y(1) \in c_l,\ldots,Y(K) \in c_i | S(1) \in h_r,\ldots,S(K) \in h_j]\\[2mm]
&= \frac{\text{\emph{Pr}}[Y(1) \in c_l,\ldots,Y(K) \in c_i , S(1) \in h_r,\ldots,S(K) \in h_j]}{\text{\emph{Pr}}[S(1) \in h_r,\ldots,S(K) \in h_j]},
\end{align*}\endgroup
for any set of quantization levels $\{ y_l,\ldots,y_i \}$, $l,\ldots,i \in \{1,\ldots,N_Y\}$ and $\{ h_r,\ldots,h_j \}$, $r,\ldots,j \in \{1,\ldots,N_S\}$. Therefore, by integrating the joint density $\mathcal{N}[\begin{bsmallmatrix} \tilde{C}_K  \\ \tilde{D}_K \end{bsmallmatrix}F_K \mu^X_1 + \begin{psmallmatrix} \tilde{C}_K  \\ \tilde{D}_K \end{psmallmatrix} L_K U^{K-1},\Sigma^{Y,S}_K]$ of $((Y^K)^\top,(S^K)^\top)^\top$ over the quantization cells, we can compute the probability mass functions $p(\tilde{y}^K|\tilde{s}^K)$ and $p(\tilde{s}^K)$.
\end{remark}

In what follows, we pose the nonlinear program for solving Problem 1.

\begin{theorem}\label{th3}
Given the system dynamics \eqref{1}, sensor quantizer \eqref{quantizer}, private output quantizer \eqref{quantizerX}, probabilistic mapping \eqref{Mapping1mod}-\eqref{Mapping3mod}, time horizon $K \in \Nat$, desired distortion level $\epsilon_K \in \Real_{\geq 0}$, joint probability density $\mathcal{N}[\begin{bsmallmatrix} \tilde{C}_K  \\ \tilde{D}_K \end{bsmallmatrix}F_K \mu^X_1 + \begin{psmallmatrix} \tilde{C}_K  \\ \tilde{D}_K \end{psmallmatrix} L_K U^{K-1},\Sigma^{Y,S}_K]$ of $((Y^K)^\top,(S^K)^\top)^\top$ (given in Lemma 3), and corresponding probability mass functions $p(\tilde{y}^K|\tilde{s}^K)$ and $p(\tilde{s}^K)$, the probability mass function $p(v^K)$ that minimizes $I[\tilde{S}^K;Z^K]$ subject to the distortion constraint, $E[||Z^K - \tilde{Y}^K||^2] \leq \epsilon_K$, can be found by solving the convex program in \eqref{eq:convex_optimization15}.\\[1mm]
\textbf{Proof:} \emph{The expression for the cost and its convexity follow from Lemma \ref{lem6}. The distortion constraint follows from Lemma \ref{lem3}, and the fact that, because $\tilde{S}^K$ and $Z^K$ are conditionally independent given $\tilde{Y}^K$, the joint distribution $p(z^K,\tilde{y}^K)$ can be written as $\sum_{\tilde{s}^K \in \mathcal{S}^K} p(\tilde{s}^K)p(\tilde{y}^K|\tilde{s}^K)p(z^K|\tilde{y}^K)$. \hfill $\blacksquare$
}\end{theorem}

{The input data for the optimization problem in \eqref{eq:convex_optimization15} are: the upper bound on the allowable distortion $\epsilon_K$, the marginal probability mass function $p(\tilde{s}^K)$, and the transition probabilities $p(\tilde{y}^K|\tilde{s}^K)$. To obtain these probabilities, we need to integrate the density of $S^K$ and the joint density of $(S^K,Y^K)$ (if they are not degenerate) over the quantization cells. In Lemma 3 and Corollary 1, we provide closed form expressions for these densities and prove that they are not degenerate. Once we have these densities, they can be integrated in the sense of Remark 6 to obtain the required $p(\tilde{s}^K)$ and $p(\tilde{y}^K|\tilde{s}^K)$ and then solve problem \eqref{eq:convex_optimization15}.}

Once we have the optimal joint distribution $q^*(v^K)$ solution to \eqref{eq:convex_optimization15}, we sample from this distribution to obtain an optimal sequence of realizations $V^K = (v^*(1),\ldots,v^*(K))^\top$ that we induce to system \eqref{1} at every time-step, see Figure \ref{Fig2} and Figure \ref{Map}. We remark that these realizations are computed a priori, i.e., before we start running the system. Real-time realizations of $Y(K)$ are not needed to compute the optimal distorting distribution. We only require the probability distributions of the processes driving the system dynamics to cast \eqref{eq:convex_optimization15} and compute $q^*(v^K)$.

\begin{remark}
{Note that, even though working with $N_V^{K}$ variables is much more manageable than working with the original $(N_Y)^{2K}$ variables (see Remark \ref{largeopt1}), having $N_V^{K}$ variables could still lead to large scale optimization problems as sufficiently large $N_V$ is required to have meaningful results. Note, however, that \eqref{eq:convex_optimization15} is solved off-line, i.e., because real-time sensor realizations are not required to perform the optimization--only the distributions of the processes driving the system are needed--the optimal mappings can be computed before we start running the system. Moreover, problem \eqref{eq:convex_optimization15} is convex and real-valued. It has a smooth cost function, bounded variables, and linear constraints. There exists many algorithms in the literature to solve this class of large-scale problems efficiently and reasonably fast, see, for instance, \emph{\cite{Boyd2004,NoceWrig06,Bertsekas/99}}. }
\end{remark}

\subsection{Receding Horizon}

For the configuration given in Theorem \ref{th3}, we have addressed the problem of designing optimal probabilistic mappings to maximize privacy for a finite time horizon $K$. Infinite horizon sub-optimal mappings can be designed by repeatedly solving finite time problems of the form \eqref{eq:convex_optimization15} in a receding horizon fashion. With receding horizon we mean the following: for each time step, starting at the current time $k$, we solve a finite horizon problem of the form \eqref{eq:convex_optimization15} over a fixed horizon $K$, i.e., over the time window $\{k,k+1,\ldots,k+K-1 \}$. We sample from the obtained optimal probability distribution $q^*(v_k^{k+K-1})$ to obtain an optimal sequence of realizations $V_k^{k+K-1} = (v_k^*(1),\ldots,v_k^*(K))^\top$. We only apply the first realization, $V(k) = v_k^*(1)$, to the system and discard the rest. Then, for the next time step, $k+1$, we solve a new finite horizon problem over $\{k+1,k+2,\ldots,k+K \}$, obtain a new optimal distribution, sample from it, and again only apply the first optimal realization, $V(k+1) = v_{k+1}^*(1)$, and discard the rest. We continue shifting the prediction horizon forward to compute the sequence of receding horizon optimal probability distributions and realizations.

Note that, besides the desired distortion constant $\epsilon_K$, problem \eqref{eq:convex_optimization15} only requires the probability mass functions $p(\tilde{y}^K|\tilde{s}^K)$ and $p(\tilde{s}^K)$ of $\tilde{Y}^K|\tilde{S}^K$ and $\tilde{S}^K$, respectively, as input information. Similarly, in a receding horizon formulation, for given horizon length $K$, we only need $\epsilon_K$, and the probability distributions $p(\tilde{y}_k^{k+K-1}|s_k^{k+K-1})$ and $p(s_k^{k+K-1})$, of $\tilde{Y}_k^{k+K-1}|S_k^{k+K-1}$ and $S_k^{k+K-1}$, respectively, to cast the corresponding optimization problem, where $\tilde{Y}_k^{k+K-1} =  (\tilde{Y}(k)^{\top},\ldots,\tilde{Y}(k+K-1)^{\top})^{\top} \in \mathcal{Y}^{K}$ and $\tilde{S}_k^{k+K-1} =  (\tilde{S}(k)^{\top},\ldots,\tilde{S}(k+K-1)^{\top})^{\top}  \in \mathcal{S}^{K}$. To obtain these distributions, we need the density of $S_k^{k+K-1} =  (S(k)^{\top},\ldots,S(k+K-1)^{\top})^{\top}$, and the joint density of $Y_k^{k+K-1} =  (Y(k)^{\top},\ldots,Y(k+K-1)^{\top})^{\top}$ and $S_k^{k+K-1}$, and then integrate them over the quantization cells. By lifting the system dynamics \eqref{1} over $\{k,k+1,\ldots,k+K-1\}$, we can write the stacked vector $((Y_k^{k+K-1})^\top,(S_k^{k+K-1})^\top)^\top \in \Real^{K(n_s + n_y)}$, in terms of the current state $X(k)$, as follows
\begin{align}\label{stackedXYk}
\begin{bmatrix} Y_k^{k+K-1} \\ S_k^{k+K-1} \end{bmatrix} &= \begin{bmatrix} \tilde{C}_{K} \\ \tilde{D}_{K} \end{bmatrix} F_{K}X(k) + \begin{bmatrix} \tilde{C}_{K} \\ \tilde{D}_{K} \end{bmatrix}T_K M^{k+K-2}_k\\ &\hspace{5mm}+ \begin{bmatrix} \tilde{C}_{K} \\ \tilde{D}_{K} \end{bmatrix} L_K U^{k+K-2}_k + \begin{bmatrix} I \\ \mathbf{0} \end{bmatrix}W^{k+K-1}_k, \notag
\end{align}
with $L_K = T_K(I_{K-1} \otimes B)$, $\tilde{C}_K = I_K \otimes C$, $\tilde{D}_K = I_K \otimes D$, and $F_{K}$ and $T_{K}$ as defined in \eqref{stackedY}.

\begin{lemma}\label{stackedDistk}
\[ \begin{bsmallmatrix} Y_k^{k+K-1} \\ S_k^{k+K-1}  \end{bsmallmatrix} \sim \mathcal{N}[\begin{bsmallmatrix} \tilde{C}_K  \\ \tilde{D}_K \end{bsmallmatrix}F_K \mu^X_k + \begin{bsmallmatrix} \tilde{C}_K  \\ \tilde{D}_K \end{bsmallmatrix} L_K U^{k+K-2}_k,\Sigma^{Y,S}_{k,K}],\] where $\mu^X_k =E[X(k)]=A\mu^X_{k-1} + BU(k)$, positive definite covariance matrix $\Sigma^{Y,S}_{k,K}$, and $\Sigma^{Y,S}_{k,K}$ can be written in terms of $\Sigma^X_k = E[(X(k) - \mu^X_k)(X(k) - \mu^X_k)^\top]$ as follows:
\begin{equation}\label{stackeconmatrix3k}
\left\{
\begin{aligned}
\Sigma^X_k &= A\Sigma^X_{k-1}A^\top + \Sigma^M,\\
\Sigma^{Y,S}_{k,K} &= \begin{bsmallmatrix} I \\[1mm] \mathbf{0} \end{bsmallmatrix}(I_{K} \otimes \Sigma^W)\begin{bsmallmatrix} I \\[1mm] \mathbf{0} \end{bsmallmatrix}^\top + \begin{bsmallmatrix} \tilde{C}_K  \\ \tilde{D}_K \end{bsmallmatrix} F_K \Sigma^X_k F_K^\top \begin{bsmallmatrix} \tilde{C}_K  \\ \tilde{D}_K \end{bsmallmatrix}^\top\\&\hspace{5mm} + \begin{bsmallmatrix} \tilde{C}_K  \\ \tilde{D}_K \end{bsmallmatrix}T_K (I_{K-1} \otimes \Sigma^M)T_K^\top \begin{bsmallmatrix} \tilde{C}_K  \\ \tilde{D}_K \end{bsmallmatrix}^\top.
\end{aligned}
\right.
\end{equation}
\emph{The proof of Lemma \ref{stackedDistk} follows similar lines as the proof of Lemma \ref{stackedDist} and it is omitted here.}
\end{lemma}

To obtain the density of $S_k^{k+K-1}$, we marginalize the joint density of $Y_k^{k+K-1}$ and $S_k^{k+K-1}$ in Lemma \ref{stackedDistk} over $Y_k^{k+K-1}$ \cite{Ross}.

\begin{corollary}
\[ S_k^{k+K-1} \sim \mathcal{N}[\tilde{D}_KF_K \mu^X_k + \tilde{D}_K L_KU^{k+K-2}_k,\Sigma^S_{k,K}], \] with $\Sigma^S_{k,K} := \begin{bsmallmatrix} \mathbf{0} \\ I_{Kn_s}  \end{bsmallmatrix}^\top \Sigma^{Y,S}_{k,K} \begin{bsmallmatrix} \mathbf{0} \\ I_{Kn_s}   \end{bsmallmatrix}$.
\end{corollary}

Having the joint density of $Y_k^{k+K-1}$ and $S_k^{k+K-1}$ allows us to compute, for any set of quantization cells, in the sense of Remark \ref{jointXY} for $S^{K}$ and $Y^{K}$, the mass functions $p(\tilde{s}_k^{k+K-1})$ and $p(\tilde{y}_k^{k+K-1}|\tilde{s}_k^{k+K-1})$ by numerically integrating this density over the quantization cells. Hereafter, we assume that $p(\tilde{s}_k^{k+K-1})$ and $p(\tilde{y}_k^{k+K-1}|\tilde{s}_k^{k+K-1})$ are known. Following a similar reasoning as in Lemma \ref{lem6}, the cost $I[\tilde{S}^{k+K-1}_k;Z^{k+K-1}_k]$ can be written in terms of the optimization variables, $q(v_k^{k+K-1})$, and it can be proved to be convex. Moreover, following the steps in the proof of Lemma \ref{lem3}, it can be proved that the distortion metric $E[||Z^{k+K-1}_{k} - \tilde{Y}^{k+K-1}_{k}||^2]$ is linear in $p(z_k^{k+K-1}|\tilde{y}_k^{k+K-1})$ for given $p(\tilde{s}_k^{k+K-1})$ and $p(\tilde{y}_k^{k+K-1}|\tilde{s}_k^{k+K-1})$. The conditional distribution $p(z_k^{k+K-1}|\tilde{y}_k^{k+K-1})$ is a linear function of $q(v_k^{k+K-1})$, and can be written as $p(z^{k+K-1}_{k}|\tilde{y}_k^{k+K-1}) = q\big( (\bar{\alpha}(z^{k+K-1}_{k}) - \bar{\alpha}(\tilde{y}_k^{k+K-1}))\modd N_Y \big)$. Therefore, minimizing the mutual information $I[\tilde{S}^{k+K-1}_k;Z^{k+K-1}_k]$, using $q(v_k^{k+K-1})$ as optimization variables, subject to $E[||Z^{k+K-1}_{k} - \tilde{Y}^{k+K-1}_{k}||^2] \leq \epsilon_K$ is a convex program. In what follows, we give the proposed receding horizon scheme.

\noindent\rule{\hsize}{1pt}\vspace{.2mm}
\textbf{Receding Horizon Scheme}\\[1mm]
\textbf{Input: } $k,K \in \Nat$, $p(\tilde{y}_k^{k+K-1}|\tilde{s}_k^{k+K-1}) = \text{Pr}[\tilde{Y}_k^{k+K-1} = \tilde{y}_k^{k+K-1} | \tilde{S}_k^{k+K-1} = \tilde{s}_k^{k+K-1}]$, $\tilde{y}_k^{k+K-1} \in \mathcal{Y}^K$, $\tilde{s}_k^{k+K-1} \in \mathcal{S}^K$, $p(s_k^{k+K-1}) = \text{Pr}[\tilde{S}_k^{k+K-1} = \tilde{s}_k^{k+K-1}]$, $\tilde{s}_k^{k+K-1} \in \mathcal{S}^K$, and $\epsilon_K \in \Real_{>0}$.\vspace{1mm}
\begin{equation} \label{eq:convex_optimization10}
\left\{\begin{aligned}
	&q^*(\tilde{v}_k^{k+K-1}) := \argmin_{q(v_k^{k+K-1})}\ I[\tilde{S}^{k+K-1}_{k};\tilde{Y}^{k+K-1}_{k}],\\
    &\text{\emph{\emph{s.t.} }} E[||Z^{k+K-1}_{k} - \tilde{Y}^{k+K-1}_{k}||^2] \leq \epsilon_K,\\
    &p(z^{k+K-1}_{k}|\tilde{y}_k^{k+K-1})\\ &\hspace{11mm}= q\big( (\bar{\alpha}(z^{k+K-1}_{k}) - \bar{\alpha}(\tilde{y}_k^{k+K-1}))\modd N_Y \big),\\[1mm]
    &\text{and } q\left(v^{k+K-1}_{k}\right) \in \text{ Simplex}.
\end{aligned}\right.
\end{equation}\\[1mm]
\textbf{Output:} $q^*(\tilde{v}_k^{k+K-1})$.\\
\vspace{.2mm}\noindent\rule{\hsize}{1pt}\vspace{1mm}

Summarizing, for given $k$, horizon $K$, and $p(\tilde{y}_k^{k+K-1}|s_k^{k+K-1})$ and $p(x_k^{k+K-1})$ corresponding to the joint density of $(Y_k^{k+K-1},S_k^{k+K-1})$ in Lemma \ref{stackedDistk}, solve problem \eqref{eq:convex_optimization10} and denote the corresponding solution as $q^*\left(v^{k+K-1}_{k}\right)$. This optimal distribution is the output of our receding scheme at time $k$. Once the optimal distributions have been computed, we sample from $q^*\left(v^{k+K-1}_{k}\right)$ to obtain an optimal sequence of realizations $V_k^{k+K-1} = (v_k^*(1),\ldots,v_k^*(K))^\top$, and only apply the first realization, $V(k) = v_k^*(1)$, to the system at time $k$, and discard the rest. Then, at time $k+1$, we sample from $q^*\left(v^{k+K}_{k+1}\right)$ and only apply the first realization, $V(k+1) = v_{k+1}^*(1)$, and discard the rest. We repeat the procedure for increasing $k$.

\begin{remark}\label{heuristic_formulation}
{Note that the distorting mechanisms resulting from the receding horizon scheme in \eqref{eq:convex_optimization10} are not optimal for the complete trajectory (although they are optimal for every subinterval). This is a limitation for most applications that use receding horizon formulations (e.g., Model Predictive Control (MPC) schemes suffer from the same heuristic nature -- MPC is in general not optimal). Optimality is only guaranteed for the finite horizon results in Section \label{finite_horizon}. We hint at how to use these results in an infinite horizon fashion using the receding horizon scheme proposed above but, again, the resulting distorting mechanisms are not guaranteed to be optimal.}
\end{remark}

\section{Simulation Experiments}

We illustrate the performance of our tools through a case study of a well stirred chemical reactor with heat exchanger. This case study has been developed over the years as a benchmark example for control systems and fault detection, see, e.g., \cite{Patton_Book}-\nocite{Wata}\cite{IET_CARLOS_JUSTIN} and references therein. The state, inputs, and output of the reactor are:
\begingroup\makeatletter\def\f@size{9.0}\check@mathfonts
\def\maketag@@@#1{\hbox{\m@th\normalsize\normalfont#1}}%
\begin{align*}
\left\{
\begin{array}{ll}
X(t) =  \begin{pmatrix} C_0\\T_0\\T_w\\T_m \end{pmatrix},
U(t) =  \begin{pmatrix} C_u\\T_u\\T_{w,u} \end{pmatrix},
Y(t) =  T_0,
\end{array}
\right.
\end{align*}\endgroup
where
\begingroup\makeatletter\def\f@size{9.0}\check@mathfonts
\def\maketag@@@#1{\hbox{\m@th\normalsize\normalfont#1}}%
\begin{align*}
\left\{
\begin{array}{ll}
C_0&: \text{Concentration of the chemical product},\\
T_0&: \text{Temperature of the product},\\
T_w&: \text{Temperature of the jacket water of heat exchanger},\\
T_m&: \text{Coolant temperature},\\
C_u&: \text{Inlet concentration of reactant},\\
T_u&: \text{Inlet temperature},\\
T_{w,u}&: \text{Coolant water inlet temperature}.
\end{array}
\right.
\end{align*}\endgroup
We use the discretized dynamics of the reactor introduced in \cite{IET_CARLOS_JUSTIN}. The discrete-time dynamics is of the form \eqref{1} with matrices $A,B,C$ as follows
\begingroup\makeatletter\def\f@size{9.0}\check@mathfonts
\def\maketag@@@#1{\hbox{\m@th\normalsize\normalfont#1}}%
\begin{align}\label{Simul1b}
\left\{
\begin{array}{c}
  A = \begin{bmatrix}
    0.8353 & 0      & 0      & 0     \\
    0      & 0.8324 & 0      & 0.0031\\
    0      & 0.0001 & 0.1633 & 0     \\
    0      & 0.0001 & 0.1633 & 0
\end{bmatrix}\\[8mm]
  (B|C^\top) = \begin{bmatrix}[ccc|c]
    0.0458 & 0      & 0      & 1\\
    0      & 0.0457 & 0      & 1\\
    0      &      0 & 0.0231 & 0\\
    0      & 0.0007 & 0.0006 & 0
\end{bmatrix}
\end{array}
\right.
\end{align}\endgroup
The original model in \cite{Patton_Book} does not consider sensor/system noise and reference signals, we include some arbitrary noise and references for our simulation experiments. We consider system and sensor noise with covariance matrices $\Sigma^M = \text{diag}[0.1,0.2,0.3,0.4]$ and $\Sigma^W = 0.1$, respectively, normally distributed initial condition $X(1) \sim \mathcal{N}[(6.94;13.76;1;1)^\top,I_4]$, and reference signal $U(k) = (50\cos[0.5k]^2,50\tanh[3k],-70\sin[0.1k])^\top$. As private output, we use the concentration of the chemical product; then, the matrix $D$ in \eqref{1} is given by the full row rank matrix $D = (1,0,0,0)$. The output of the system is the temperature of the product $T_0$, which could be monitored, e.g., for quality/safety reasons. Then, the aim of the privacy scheme is to hide the concentration of the reactant as much as possible without distorting temperature measurements excessively.

\begin{figure}[!htb]
  \centering
  \includegraphics[scale=.175]{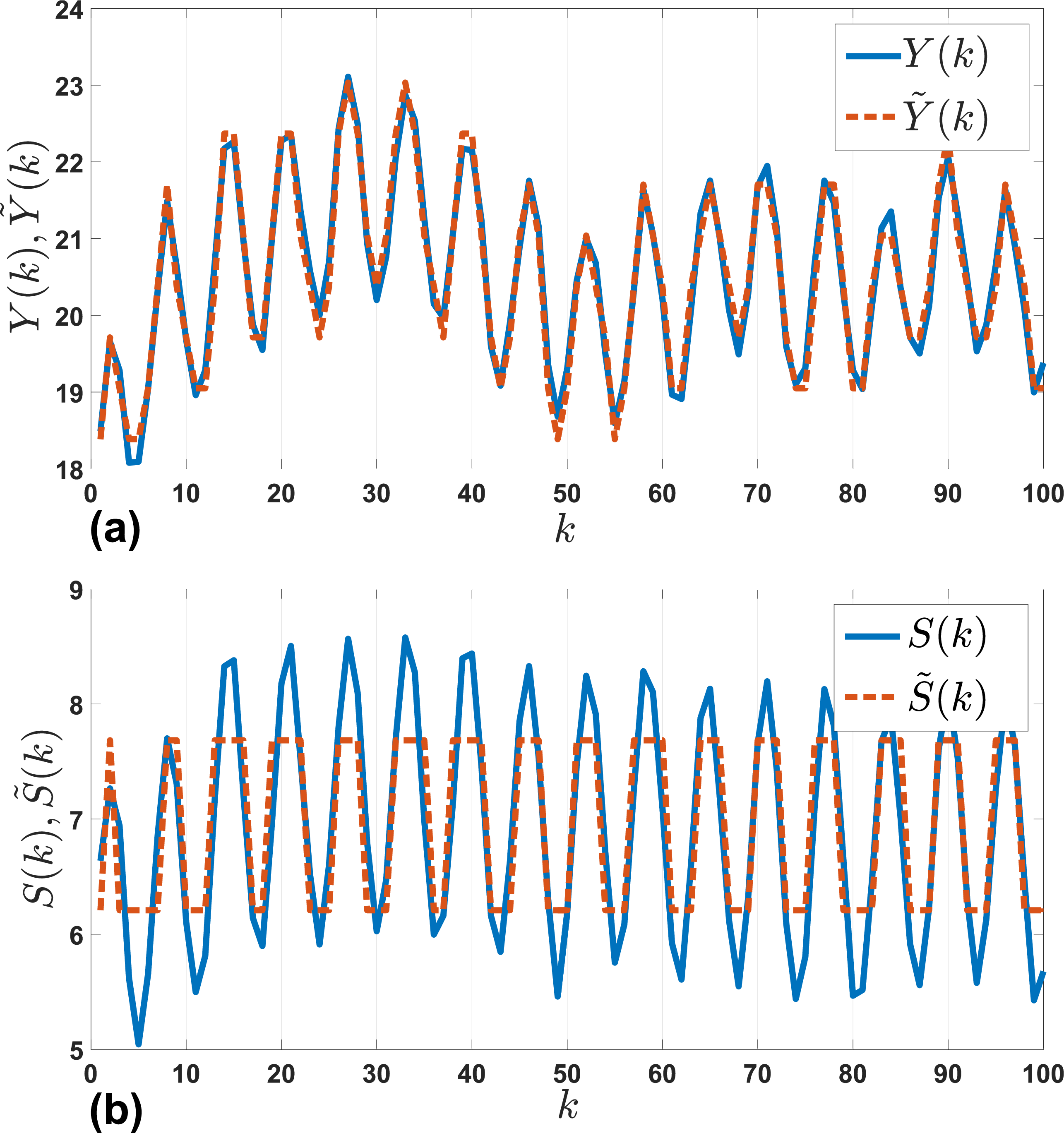}
  \caption{(a) Sample path of the output $Y(k)$ and its quantized version $\tilde{Y}(k)$ (3-bit quantizer); and (b) Sample path of the private output $S(k)$ and its quantized version $\tilde{S}(k)$ (1-bit quantizer).}\label{Simul_10}
\end{figure}

\begin{figure}[!htb]
  \centering
  \includegraphics[scale=.175]{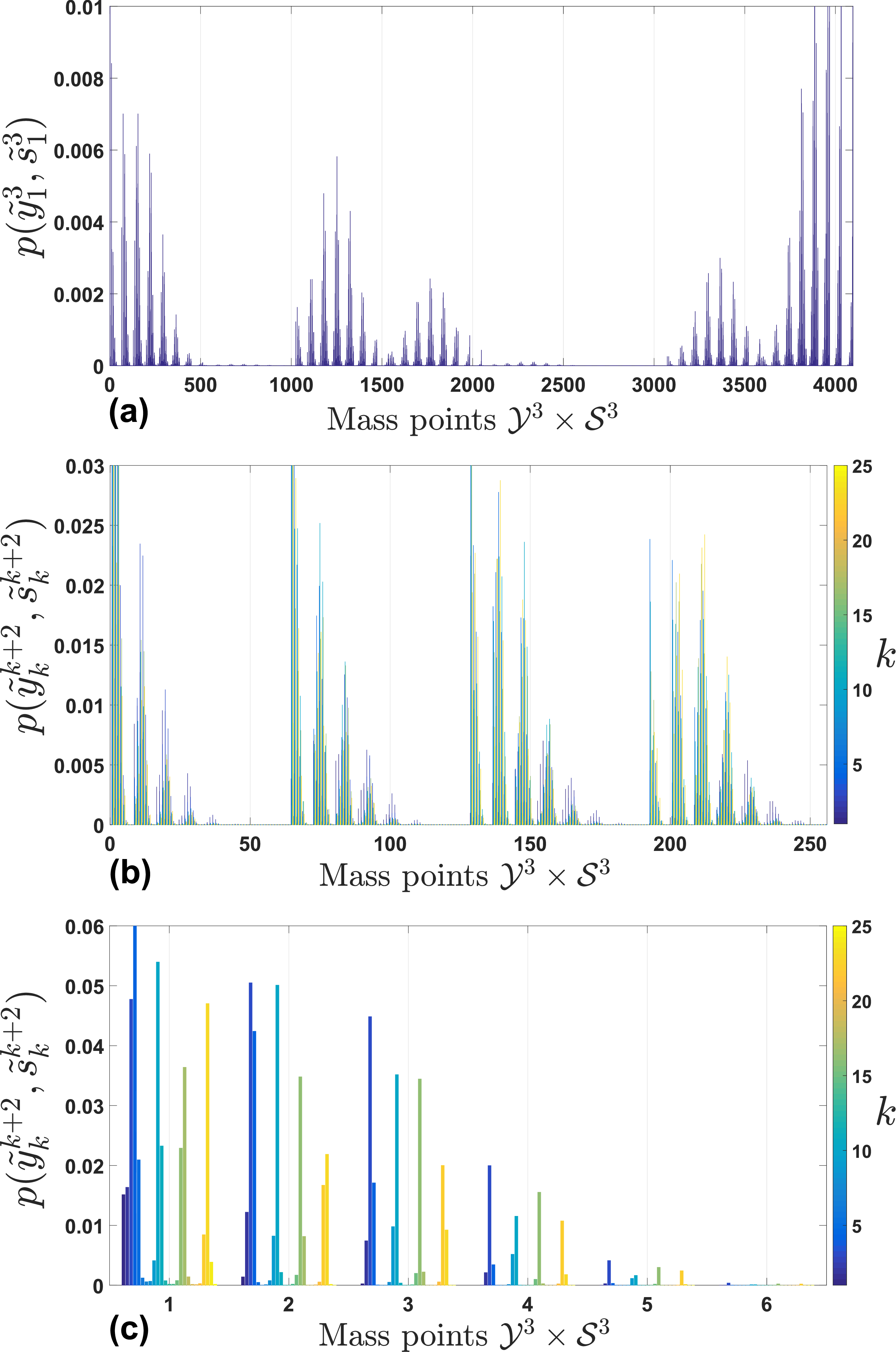}
  \caption{(a) Joint pmf $p(\tilde{y}_1^{3},\tilde{s}_1^{3})$, $\tilde{y}_1^{3},\tilde{s}_1^{3} \in \mathcal{Y}^3 \times \mathcal{S}^3$; (b) Joint pmf $p(\tilde{y}_k^{k+2},\tilde{s}_k^{k+2})$, $k=1,\ldots,25$, for the first $256$ mass points; and (c) Zoom of the probabilities of the first six mass points for increasing $k$ .}\label{Simul_11}
\end{figure}

We first consider the receding horizon formulation of Problem 1, i.e., minimizing $I[\tilde{S}^{k+K-1}_{k};Z^{k+K-1}_{k}]$ for some quantized version $\tilde{S}(k)$ of the private output $S(k)$, over sliding windows of the form $\{k,k+1,\ldots,k+K-1\}$, $k \in \Nat$. For these experiments, we use a 3-bit quantizer for $Y(k)$ ($N_Y = 8$) with levels
\begingroup\makeatletter\def\f@size{9.5}\check@mathfonts
\def\maketag@@@#1{\hbox{\m@th\normalsize\normalfont#1}}%
\begin{align*}
\mathcal{Y} &= \{y_1,\ldots,y_8\} \\ &=  \{18.38,19.04,19.71,20.37,21.04,21.70,22.36,23.03\},
\end{align*}\endgroup
and corresponding quantization cells
\begingroup\makeatletter\def\f@size{9.5}\check@mathfonts
\def\maketag@@@#1{\hbox{\m@th\normalsize\normalfont#1}}%
\begin{align*}
&\mathcal{C} =  \{(-\infty,18.71],(18.71,19.38],(19.38,20.04],(20.04,20.70],\\
&(20.70,21.37],(21.37,22.03],(22.03,22.70],(22.70,\infty)\}.
\end{align*}\endgroup
For the private output $S(k)$, we use a binary quantizer ($N_S = 2$) with levels $\mathcal{S} = \{s_1,s_2 \} = \{6.20,7.68\}$, and cells $\mathcal{H} = \{h_1,h_2\} = \{(-\infty,6.94],(6.94,\infty)\}$. These quantizers were selected based on system trajectories to avoid having always saturated $\tilde{Y}(k)$ and $\tilde{S}(k)$, see Figure \ref{Simul_10}. Then, for horizon $K=3$ and $k \in \Nat$, the stacked quantized output $\tilde{Y}_{k}^{k+K-1}$ (and thus also $Z_{k}^{k+K-1}$) and the stacked private output $\tilde{S}_{k}^{k+K-1}$ have alphabets $\mathcal{Y}^3$ (with $(N_Y)^K = 512$ elements) and $\mathcal{S}^3$ (with $(N_S)^K = 8$ elements) given by
\begin{align}
\mathcal{Y}^3 &= \small \left\{ \begin{bmatrix} y_1 \\ y_1 \\ y_1\end{bmatrix}, \begin{bmatrix} y_2 \\ y_1 \\ y_1\end{bmatrix},
\begin{bmatrix} y_3 \\ y_1 \\ y_1 \end{bmatrix},\ldots,
\begin{bmatrix} y_6 \\ y_8 \\ y_8 \end{bmatrix},
\begin{bmatrix} y_7 \\ y_8 \\ y_8 \end{bmatrix},
\begin{bmatrix} y_8 \\ y_8 \\ y_8 \end{bmatrix}
\normalsize \right\}, \label{masspoints21}\\
\mathcal{S}^3 &= \small \left\{ \begin{bmatrix} s_1 \\ s_1 \\ s_1 \end{bmatrix}, \begin{bmatrix} s_2 \\ s_1 \\ s_1 \end{bmatrix},
\begin{bmatrix} s_1 \\ s_2 \\ s_1 \end{bmatrix},
\begin{bmatrix} s_2 \\ s_2 \\ s_1 \end{bmatrix},\ldots,
\begin{bmatrix} s_1 \\ s_2 \\ s_2 \end{bmatrix},
\begin{bmatrix} s_2 \\ s_2 \\ s_2 \end{bmatrix}
\normalsize \right\}, \label{masspoints22}
\end{align}
with $y_i$, $i=1,2,\ldots,8$, and $s_j$, $j \in \{1,2\}$, as introduced above. We let the distorting random process $V(k)$ (see \eqref{Mapping1mod}-\eqref{Mapping3mod}) have an alphabet with $N_V = 5$ elements, i.e., $\mathcal{V} = \{0,1,2,3,4\}$. Then, for $K=3$, the stacked vector $V_{k}^{k+2}$ has alphabet $\mathcal{V}^3$ with $(N_V)^3 = 125$ elements. To cast problem \eqref{eq:convex_optimization10}, we need the probability mass functions $p(\tilde{y}_k^{k+2}|\tilde{s}_k^{k+2}) = p(\tilde{y}_k^{k+2},\tilde{s}_k^{k+2})/p(\tilde{s}_k^{k+2})$ and $p(\tilde{s}_k^{k+2}) = \sum_{\tilde{s}_k^{k+2} \in \mathcal{S}^3} p(\tilde{y}_k^{k+2},\tilde{s}_k^{k+2})$, $\tilde{y}_k^{k+2} \in \mathcal{Y}^3$, $\tilde{s}_k^{k+2} \in \mathcal{S}^3$. These distributions are fully characterized by the joint pmf, $p(\tilde{y}_k^{k+2},\tilde{s}_k^{k+2})$, $\tilde{y}_k^{k+2},\tilde{s}_k^{k+2} \in \mathcal{Y}^3 \times \mathcal{S}^3$, where the alphabet $\mathcal{Y}^3 \times \mathcal{S}^3$ is given by
\begingroup\makeatletter\def\f@size{9}\check@mathfonts
\def\maketag@@@#1{\hbox{\m@th\normalsize\normalfont#1}}%
\begin{align}
&\mathcal{Y}^3 \times \mathcal{S}^3 = \left\{ \begin{bmatrix} y_1 \\ y_1 \\ y_1 \\ s_1 \\ s_1 \\ s_1 \end{bmatrix}, \begin{bmatrix} y_2 \\ y_1 \\ y_1 \\ s_1 \\ s_1 \\ s_1 \end{bmatrix},\ldots,
\begin{bmatrix} y_7 \\ y_8 \\ y_8 \\ s_1 \\ s_1 \\ s_1 \end{bmatrix},
\begin{bmatrix} y_8 \\ y_8 \\ y_8 \\ s_1 \\ s_1 \\ s_1 \end{bmatrix},\right.\\
&\left.\begin{bmatrix} y_1 \\ y_1 \\ y_1 \\ s_2 \\ s_1 \\ s_1 \end{bmatrix},\ldots,
\begin{bmatrix} y_8 \\ y_8 \\ y_8 \\ s_2 \\ s_1 \\ s_1 \end{bmatrix},\ldots,
\begin{bmatrix} y_8 \\ y_8 \\ y_8 \\ s_3 \\ s_3 \\ s_3 \end{bmatrix}
\normalsize \right\}, \label{masspoints31}
\end{align}\endgroup
with $|\mathcal{Y}^3 \times \mathcal{S}^3| = (N_YN_S)^3= 4096$. We integrate (in the sense of Remark \ref{jointXY}) the joint density of $Y_k^{k+2}$ and $S_k^{k+2}$ (given in Lemma \ref{stackedDistk}) over the quantization cells, $\mathcal{C}$ and $\mathcal{H}$, to obtain $p(\tilde{y}_k^{k+2},\tilde{s}_k^{k+2})$ (and thus also $p(\tilde{s}_k^{k+2})$ and $p(\tilde{y}_k^{k+2}|\tilde{s}_k^{k+2})$) for $k \in \{1,\ldots,25\}$. In Figure \ref{Simul_11}(a), we show the joint probability mass function $p(\tilde{y}_1^{3},\tilde{s}_1^{3})$, $\tilde{y}_1^{3},\tilde{s}_1^{3} \in \mathcal{Y}^3 \times \mathcal{S}^3$. The mass points are indexed following the ordering logic in \eqref{masspoints31}. Figure \ref{Simul_11}(b) depicts the joint pmf $p(\tilde{y}_k^{k+2},\tilde{s}_k^{k+2})$, $k=1,\ldots,25$, for the first $256$ mass points of the alphabet $\mathcal{Y}^3 \times \mathcal{S}^3$ (indexed as in \eqref{masspoints31}); and, in Figure \ref{Simul_11}(c), we show zooms of the probabilities of the first six mass points for increasing $k$. Note that there is a lot of variability in the probabilities of the mass points as $k$ increases.

\begin{figure}[!htb]
  \centering
  \includegraphics[scale=.19]{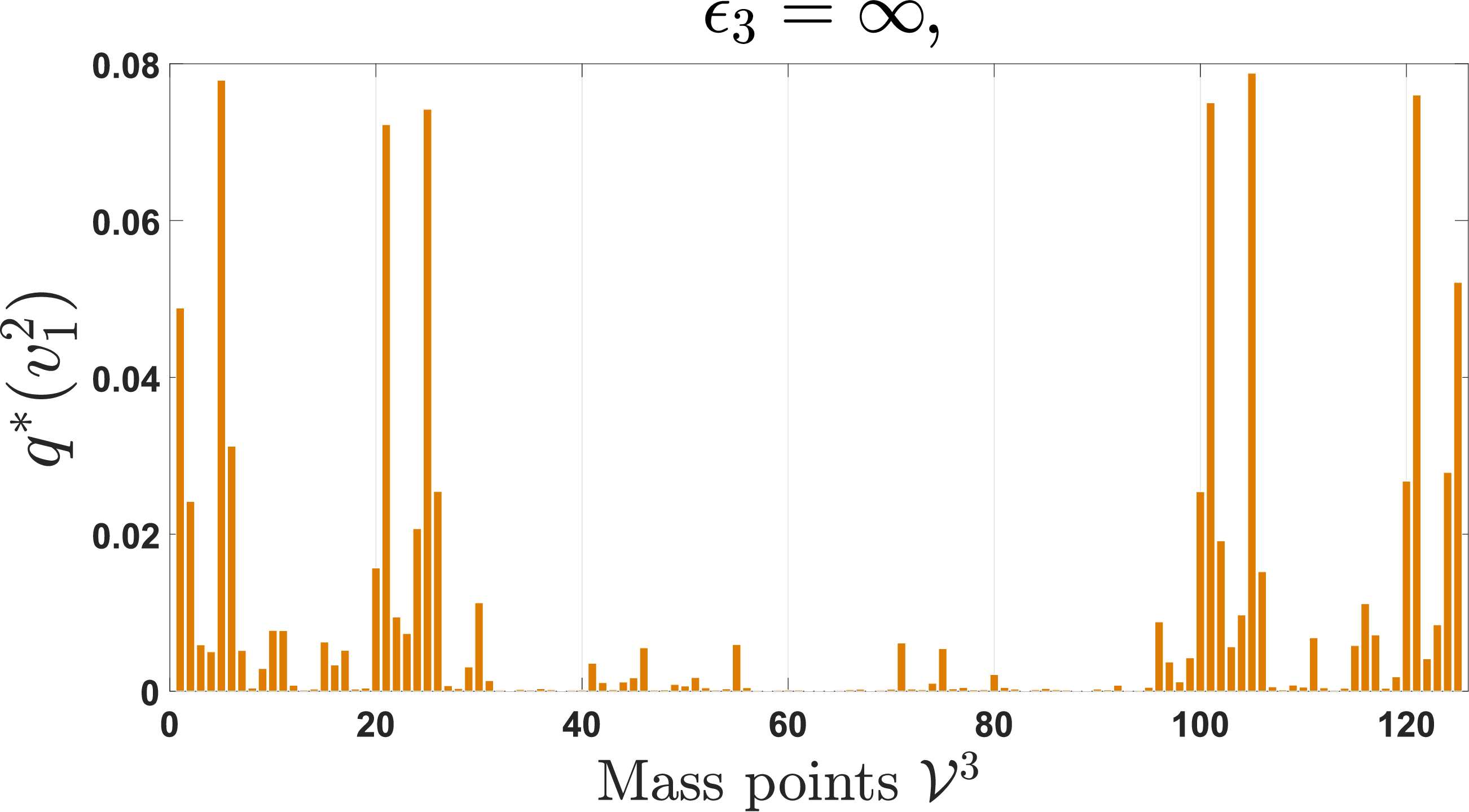}
  \caption{Optimal distorting distribution $q^*(v_1^{2})$ for $\epsilon_K = \infty$.}\label{Simul_12}
\end{figure}

\begin{figure}[!htb]
  \centering
  \includegraphics[scale=.19]{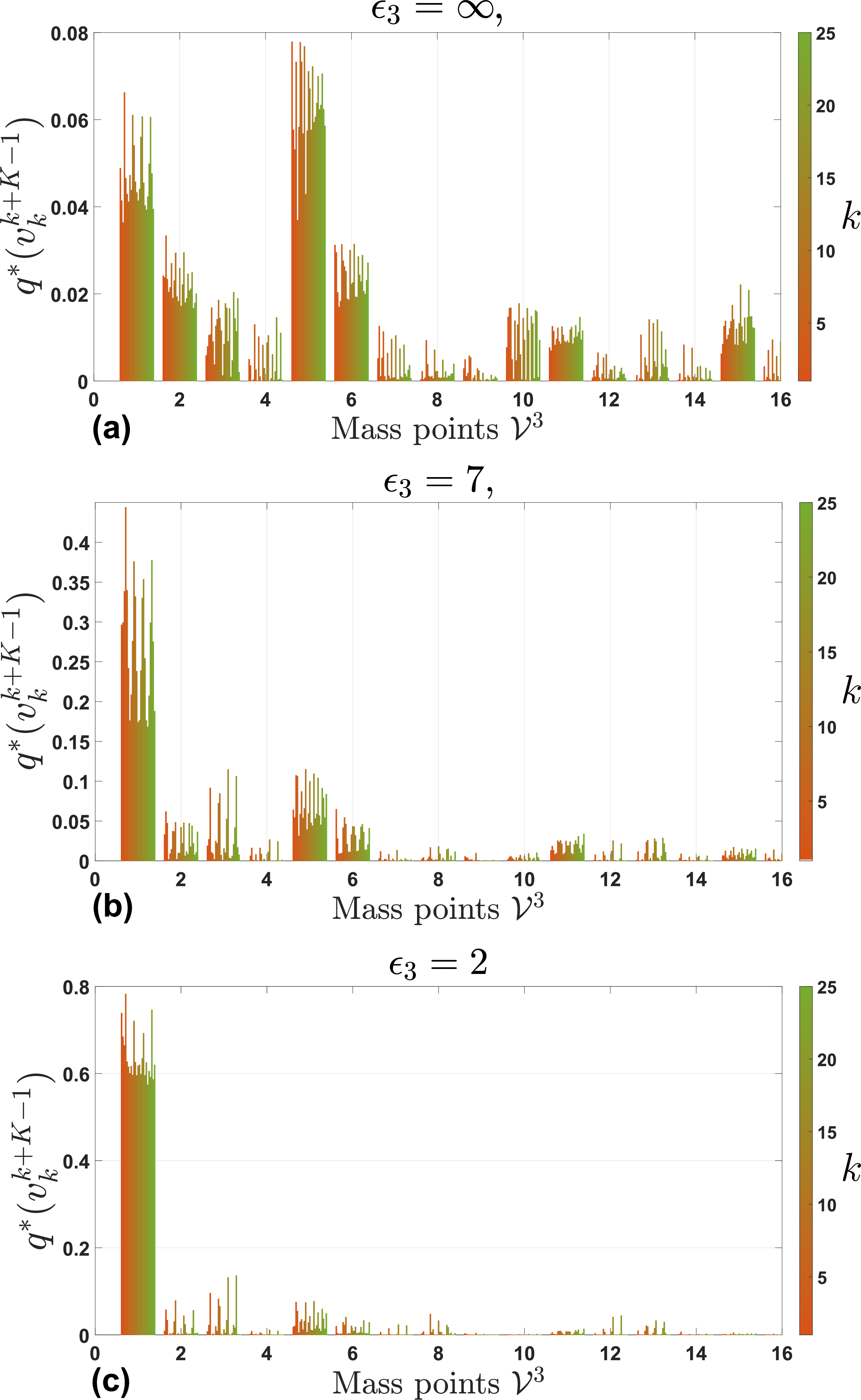}
  \caption{Evolution of the optimal probability distribution $q^*(v_k^{k+2})$ solution to \eqref{eq:convex_optimization10} for increasing $k$ and different distortion upper bounds $\epsilon_K$.}\label{Simul_13}
\end{figure}

\begin{figure}[!htb]
  \centering
  \includegraphics[scale=.19]{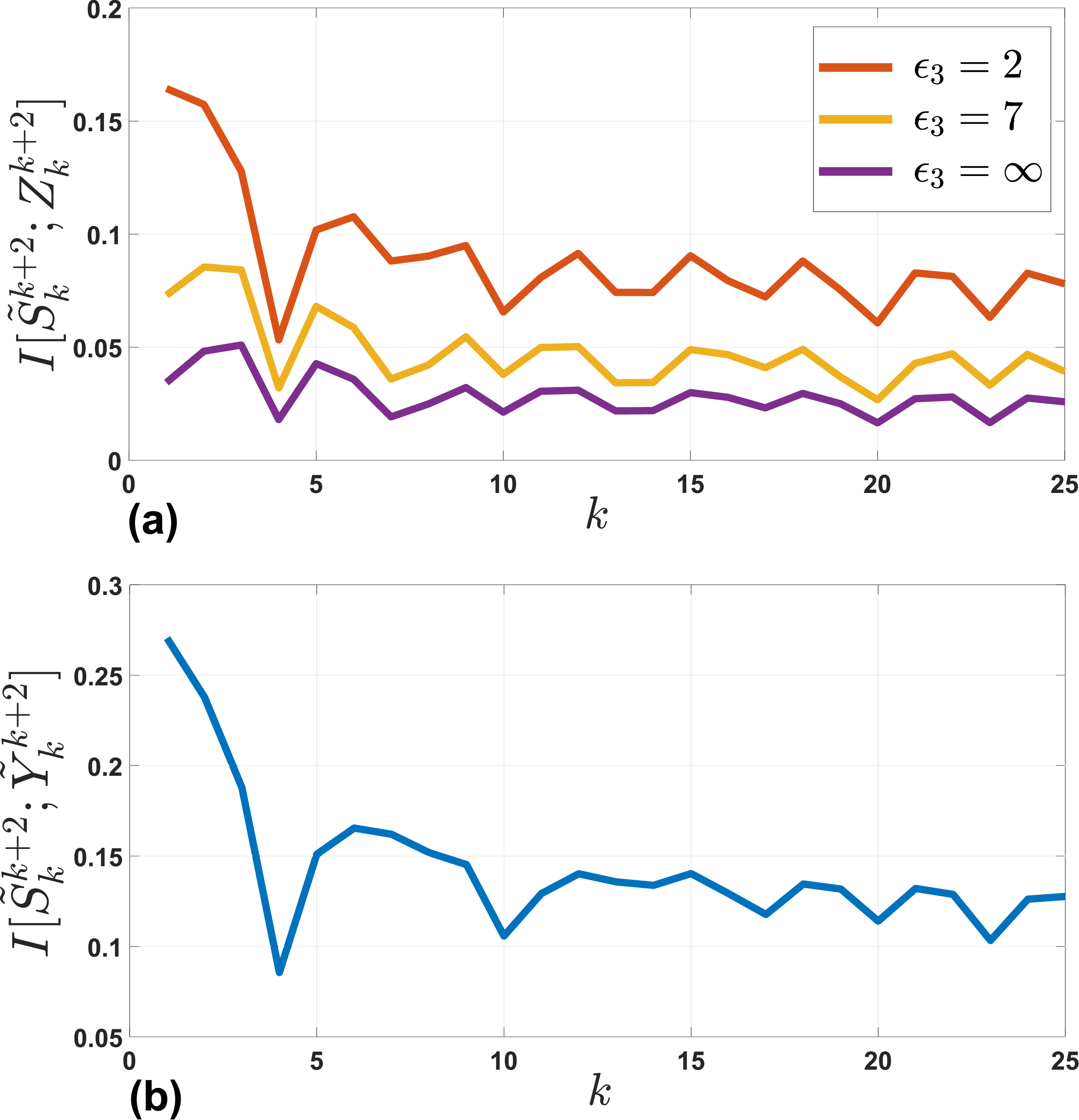}
  \caption{For $K=3$ and increasing $k$: \textbf{(a)} Evolution of the optimal cost $I[\tilde{S}_k^{k+K-1};Z_k^{k+K-1}]$; and \textbf{(b)} Mutual information between $\tilde{S}_k^{k+K-1}$ and $\tilde{Y}_k^{k+K-1}$, $I[\tilde{S}_k^{k+K-1};\tilde{Y}_k^{k+K-1}]$ (information disclosed if no privacy preserving mapping was in place).}\label{Simul_14}
\end{figure}

\begin{figure}[!htb]
  \centering
  \includegraphics[scale=.19]{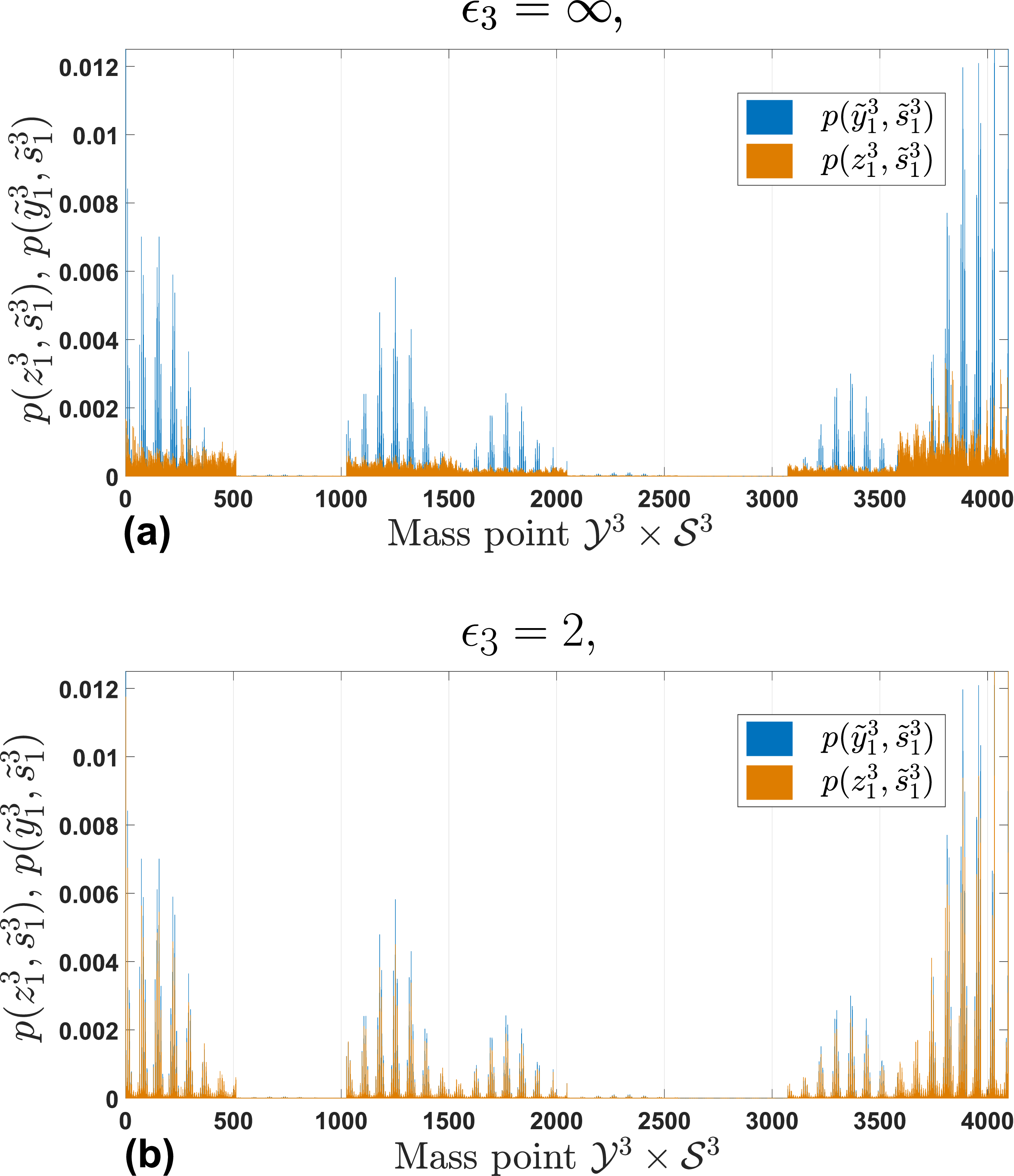}
  \caption{For $K=3$: \textbf{(a)} Joint probability distributions $p(z_1^3,\tilde{s}_1^3)$ and $p(\tilde{y}_1^3,\tilde{s}_1^3)$ for $\epsilon_K = \infty$; and \textbf{(b)} $p(z_1^3,\tilde{s}_1^3)$ and $p(\tilde{y}_1^3,\tilde{s}_1^3)$ for $\epsilon_K = 2$. Note that as $\epsilon_K \rightarrow 0$, $p(z_1^3,\tilde{s}_1^3) \rightarrow p(\tilde{y}_1^3,\tilde{s}_1^3)$.}\label{Simul_15}
\end{figure}

Next, for given $p(\tilde{y}_k^{k+2},\tilde{s}_k^{k+2})$, we show results of the receding horizon scheme \eqref{eq:convex_optimization10} for horizon $K=3$,\linebreak $k=1,2,\ldots,25$, and three different levels of distortion, $\epsilon_K = \infty,7,3$, where $\epsilon_K = \infty$ means that the optimization problems in \eqref{eq:convex_optimization10} are solved without considering the distortion constraint. In Figure \ref{Simul_12}, we show the optimal distorting distribution $q^*(v_1^{2})$ for $\epsilon_K = \infty$ and, in Figure \ref{Simul_13}, the evolution (of the first sixteen mass points) of the optimal distribution $q^*(v_k^{k+2})$ solution to \eqref{eq:convex_optimization10} for increasing $k$ and $\epsilon_K = \infty,7,2$. The mass points are indexed following the ordering logic in \eqref{masspoints31}. Note that there is a lot of probability variability both among mass points and in time (as $k$ grows). The optimal distorting distributions follow some nontrivial patterns that depend on the dynamics, quantizer, and desired distortion level. Further note that even the unconstraint formulation ($\epsilon_K = \infty$) does not lead to uniform distributions; and, as $\epsilon_K \rightarrow 0$, the sequence of optimal distributions $q^*(v_k^{k+2})$ concentrates most of its probability at the first mass point (the zero vector). This is what one would expect as the zero vector leads to no distortion. For $\epsilon_K = \infty,7,2$, Figure \ref{Simul_14} depicts the evolution of the optimal cost $I[\tilde{S}_k^{k+K-1};Z_k^{k+K-1}]$ and the mutual information $I[\tilde{S}_k^{k+K-1};\tilde{Y}_k^{k+K-1}]$. As expected, $I[\tilde{S}_k^{k+K-1};Z_k^{k+K-1}]$ decreases for increasing $\epsilon_K$ uniformly in $k$. The mutual information $I[\tilde{S}_k^{k+K-1};\tilde{Y}_k^{k+K-1}]$ in Figure \ref{Simul_14}(b) characterizes the disclosed information if no privacy preserving mapping was in place. Finally, in Figure \ref{Simul_15}, we show the joint probability distributions $p(z_1^3,\tilde{s}_1^3)$ and $p(\tilde{y}_1^3,\tilde{s}_1^3)$ for $\epsilon_K = \infty$ and $\epsilon_K = 2$. Note that, as one would expect, $p(z_1^3,\tilde{s}_1^3) \rightarrow p(\tilde{y}_1^3,\tilde{s}_1^3)$ as $\epsilon_K \rightarrow 0$. That is, as we allow for less distortion, we have less freedom to reshape $p(\tilde{y}_1^3,\tilde{s}_1^3)$ by passing $\tilde{Y}_1^3$ through $p(\tilde{y}_1^3,\tilde{z}_1^3)$ before transmission.

\begin{figure}[!htb]
  \centering
  \includegraphics[scale=.26]{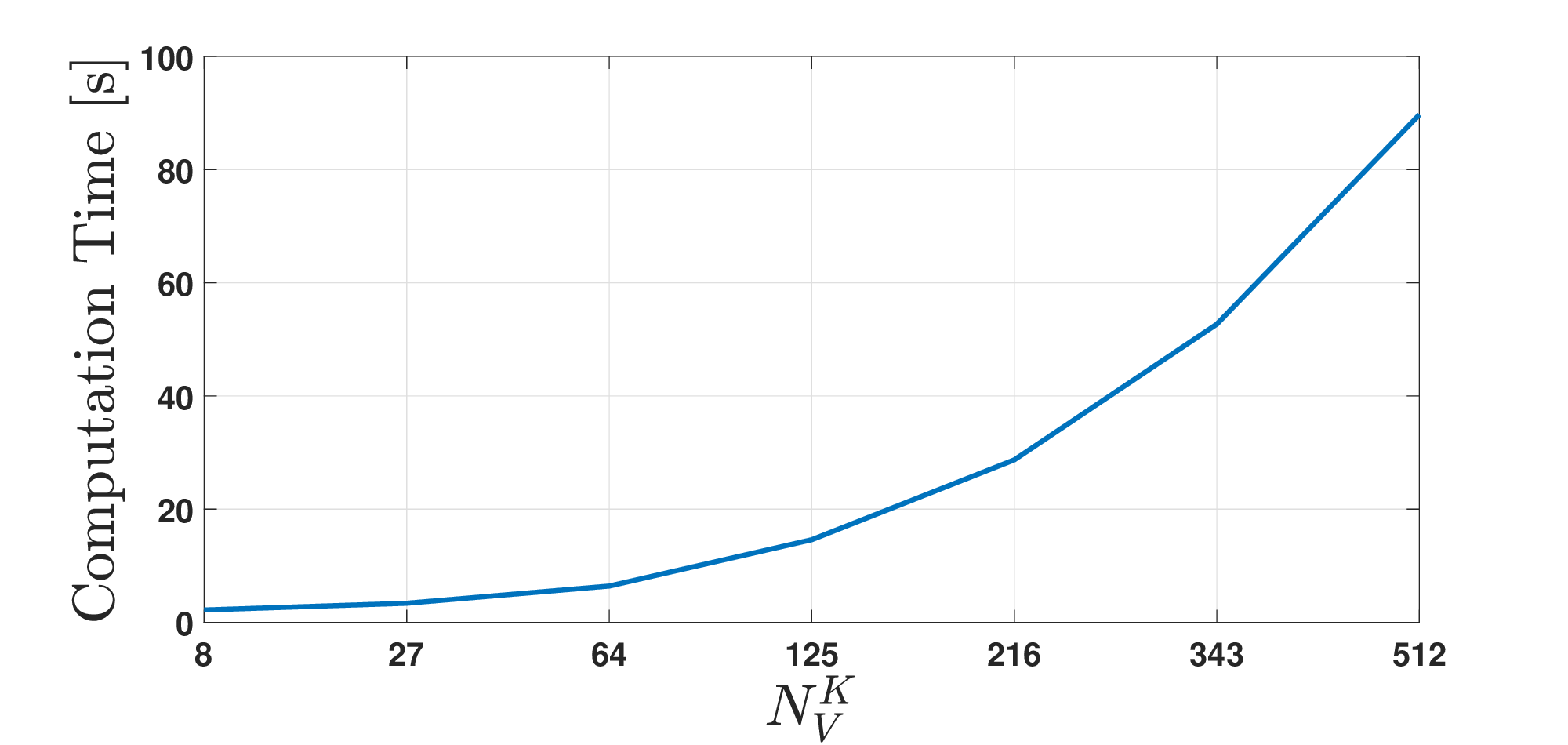}
  \caption{Computation time for problem \eqref{eq:convex_optimization10}, with $K=3$ and $k=1$, as the number of variables, $N_V^K$, increases.}\label{last}
\end{figure}

{We remark that all the above computations were performed on a PC, Intel 2.70 GHz, in Matlab 2015b (using the parallel computing toolbox with four cores). In Figure \ref{Simul_12}, we show the optimal distribution $q^*(v_1^{2})$, i.e., we solve the optimization problem in \eqref{eq:convex_optimization10} once (for $K=3$ and $k=1$) using 125 variables ($N_V = 5$) and $\epsilon_3 = \infty$ (unconstraint case). This optimization took 8.7 seconds to be performed. The same optimization but with $\epsilon_3 = 8$ (considering the distortion constraint) took 14.6 seconds. In Figure \ref{last}, for $\epsilon_3 = 8$, we show the evolution of the computation time as the cardinality $N_V = |\mathcal{V}|$ of the alphabet of the additive process $V(k)$ increases (i.e., as we increase the number of variables $N_V^K$). The computation time grows exponentially with the number of variables. Note, however, that the number of variables is independent of the size of the quantizers. We fix $|V|$ a priori and then solve \eqref{eq:convex_optimization10} for given quantizers. What changes with the size of the quantizers is the structure of the cost function as finer quantizers lead to increasingly involved expressions in \eqref{eq:convex_optimization10}. Finally, in Figure \ref{Simul_13}(a), we show the evolution of the optimal $q^*(v_k^{k+K-1})$ for $K=3$ and $k=1,2,\ldots,16$. That is, we solve the optimization problem in \eqref{eq:convex_optimization10} sixteen times following the receding horizon formulation. It took 160.5 seconds to perform these sixteen optimizations.}

\section{Conclusion}
For a class of Cyber-Physical-Systems (CPSs), we have presented a detailed mathematical framework built around systems theory, information theory, and convex optimization to deal with privacy problems raised by the use of public/unsecured communication networks to transmit sensor data. In particular, to prevent adversaries from obtaining an accurate estimate of the private part of the system state, we have provided tools (in terms of convex programs) to optimally randomize (via some probabilistic mappings) sensor data before transmission for a desired level of distortion. That is, given a maximum level of distortion tolerated by a particular application, we give tools to synthesize probabilistic mappings that maximize privacy (in the sense of hiding the private output as much as possible) while satisfying the distortion constraint on the original sensor data. Our tools are capable of dealing with the dynamic and non-stationary nature of CPSs at the price of having to solve some medium to large scale optimization problems. We have presented extensive simulation experiments to show the performance of out tools. Note that we have found some nontrivial distorting probability distributions that highly depend on the system dynamics, quantizer, and desired distortion level.

\section*{Acknowledgment}
This work was supported by the Australian Research Council (ARC) under the Project DP170104099; the NATO Science for Peace and Security (SPS) PROGRAMME under the project SPS.SFP G5479; and a Schmidt Data-X Grant from the Princeton Center for Statistics and Machine Learning.

\section*{Appendices}

\subsection{Proof of Lemma 1}

The expression on the right-hand side of \eqref{cost3} follows by inspection of Definition \ref{mutual_info}, and the fact that $p(\tilde{s}^K,z^K) = p(\tilde{s}^K)p(z^K|\tilde{s}^K)$ (chain rule). By \cite[Theorem 2.7.4]{Cover}, cost \eqref{cost3} is convex in $p(z^K|\tilde{s}^K)$ for given $p(\tilde{s}^K)$. However, our optimization variables are $q(v^K)$ and not $p(z^K|\tilde{s}^K)$. {By construction, $\tilde{S}^K$ and $Z^K$ are conditionally independent given $\tilde{Y}^K$. That is, given the information flow in the system (see Figure 1), all the information that $Z^K$ might carry about $\tilde{S}^K$ is carried by $\tilde{Y}^K$ because $Z^K = G(\tilde{Y}^K)$ is a random function of $\tilde{Y}^K$ only. The later implies that given $\tilde{Y}^K$, $\tilde{S}^K$ and $Z^K$ are conditionally independent (see, e.g., \cite[Sections 2.8 and 4.1]{Cover} for details). It follows that $p(\tilde{s}^K,\tilde{y}^K,z^K) = p(\tilde{s}^K)p(\tilde{y}^K|\tilde{s}^K)p(z^K|\tilde{y}^K)$.} Using this expression for $p(\tilde{s}^K,\tilde{y}^K,z^K)$, we can write \eqref{cost3a} and \eqref{cost3b} by conditioning and marginalizing the joint distribution. It follows that, by combining \eqref{cost3}-\eqref{cost3b}, we can write $I[\tilde{S}^K;Z^K]$ in terms of $p(z^K|\tilde{y}^K)$, $p(\tilde{y}^K|\tilde{s}^K)$, and $p(\tilde{s}^K)$, i.e., the cost is a function of $p(z^K|\tilde{y}^K)$ and it is parametrized by $p(\tilde{y}^K|\tilde{s}^K)$ and $p(\tilde{s}^K)$. Convexity with respect to $p(z^K|\tilde{y}^K)$ follows from convexity with respect to $p(z^K|\tilde{s}^K)$ because $p(z^K|\tilde{s}^K)$ is just a linear combination of $p(z^K|\tilde{y}^K)$, see \eqref{cost3a}, and convexity is preserved under affine transformations \cite{Boyd2004}. By definition, $p(z^K,\tilde{y}^K) = \text{Pr}[Z^K=z^K,\tilde{Y}^K=\tilde{y}^K]$, $z^K,\tilde{y}^K \in \mathcal{Y}^K$. Note that
\begin{align*}
&\text{Pr}[Z^K=z^K,\tilde{Y}^K=\tilde{y}^K]\\ &= \text{Pr}[Z(1)=z(1),\ldots,Z(K)=z(K),\\[1mm]
&\hspace{30mm}\tilde{Y}(1)=\tilde{y}(1),\ldots,\tilde{Y}(K)=\tilde{y}(K)],
\end{align*}
for  $z(i),\tilde{y}(j) \in \mathcal{Y}$, $i,j \in \{1,\ldots,K \}$. Using \eqref{Mapping3mod}, we can further expand $\text{Pr}[Z^K=z^K,\tilde{Y}^K=\tilde{y}^K]$ as follows
\begin{align*} &\text{Pr}[Z^K=z^K,\tilde{Y}^K=\tilde{y}^K]\\[1mm]
&=\text{Pr}[\beta\big( ( \alpha(\tilde{Y}(1))+V(1) )\modd N_Y \big)=z(1),\ldots,\\
&\hspace{30mm}\tilde{Y}(1)=\tilde{y}(1),\ldots,\tilde{Y}(K)=\tilde{y}(K)]\\[1mm]
&=\text{Pr}[V(1) = \big( \alpha(z(1)) - \alpha(\tilde{y}(1)) \big)\modd N_Y ,\ldots,\\
&\hspace{30mm}\tilde{Y}(1)=\tilde{y}(1),\ldots,\tilde{Y}(K)=\tilde{y}(K)]\\[1mm]
&\overset{\text{(a})}{=}\text{Pr}[V(1) = \big( \alpha(z(1)) - \alpha(\tilde{y}(1)) \big)\modd N_Y ,\ldots,\\
&\hspace{4mm}V(K) = \big( \alpha(z(K)) - \alpha(\tilde{y}(K)) \big)\modd N_Y ]\text{ Pr}[\tilde{Y}^K=\tilde{y}^K]\\[1mm]
&\overset{\text{(b})}{=}\text{Pr}[V^K = \big( \bar{\alpha}(z^K) - \bar{\alpha}(\tilde{y}^K) \big)\modd N_Y]\text{ Pr}[\tilde{Y}^K=\tilde{y}^K]\\[1mm]
&\overset{\text{(c})}{=} q\big( \big( \bar{\alpha}(z^K) - \bar{\alpha}(\tilde{y}^K) \big)\modd N_Y \big)p(\tilde{y}^K),
\end{align*}
where (a) follows from independence between $V^K$ and $\tilde{Y}^K$, (b) from the definition of $\bar{\alpha}(\cdot)$ in \eqref{stacked_alpha}, and (c) by construction of $q(v^K)$ since $q(v^K) = \text{Pr}[V^K = v^k]$, $v^k \in \mathcal{V}^K$. It follows that $p(z^K|\tilde{y}^K) = p(z^K,\tilde{y}^K)/p(\tilde{y}^K) = q\big( \big( \bar{\alpha}(z^K) - \bar{\alpha}(\tilde{y}^K) \big)\modd N_Y \big)$ and thus \eqref{cost3c} holds true. It remains to prove that $I[\tilde{S}^K;Z^K]$ is convex in $q(v^K)$ for given $p(\tilde{y}^K)$ and $p(\tilde{s}^K|\tilde{y}^K)$. We have concluded convexity of $I[\tilde{S}^K;Z^K]$ with respect to $p(z^K|\tilde{y}^K)$ above. Hence, because $p(z^K|\tilde{y}^K) = q\big( \big( \bar{\alpha}(z^K) - \bar{\alpha}(\tilde{y}^K) \big)\modd N_Y \big)$ and $q\big( \big( \bar{\alpha}(z^K) - \bar{\alpha}(\tilde{y}^K) \big)\modd N_Y \big)$ is a linear transformation of $q(v^K)$ (note that $q\big( \big( \bar{\alpha}(z^K) - \bar{\alpha}(\tilde{y}^K) \big)\modd N_Y \big) = q(v^K)$ for $\big( \bar{\alpha}(z^K) - \bar{\alpha}(\tilde{y}^K) \big)\modd N_Y = v^K$ and zero otherwise), the cost $I[\tilde{S}^K;Z^K]$ is convex in $q(v^K)$ because convexity is preserved under affine transformations. \hfill $\blacksquare$

\subsection{Proof of Lemma 2}

Define the function $d(Z^K,\tilde{Y}^K):= ||Z^K - \tilde{Y}^K||^2$. The function $d(Z^K,\tilde{Y}^K)$ is a deterministic function of two jointly distributed random vectors, $Z^K$ and $\tilde{Y}^K$, with joint distribution $p(z^K,\tilde{y}^K)$. The joint distribution can be written as $p(z^K,\tilde{y}^K) = p(z^K|\tilde{y}^K)p(\tilde{y}^K)$ (chain rule), and, by \eqref{cost3c}, the conditional probability distribution is given by $p(z^K|\tilde{y}^K) = q\big( (\bar{\alpha}(z^K) - \bar{\alpha}(\tilde{y}^K))\modd N_Y$ (see the proof of Lemma 1 for details). That is, $p(z^K|\tilde{y}^K)$ is a linear transformation of $q(v^K)$. Therefore, see, e.g., \cite{Ross} for details, $E[d(Z^K,\tilde{Y}^K)] = \sum_{\tilde{y}^K \in \mathcal{Y}^K}\sum_{z^K \in \mathcal{Y}^K}p(z^K,\tilde{y}^K)d(z^K,\tilde{y}^K)$, and because the joint distribution can be written as $p(z^K,\tilde{y}^K) = q\big( (\bar{\alpha}(z^K) - \bar{\alpha}(\tilde{y}^K))\modd N_Y \big)p(\tilde{y}^K)$, the expected distortion $E[d(Z^K,\tilde{Y}^K)]$ is given by \eqref{distortion}, and it is linear in $q(v^K)$ for given $p(\tilde{y}^K)$. \hfill $\blacksquare$

\subsection{Proof of Lemma 3}

To simplify notation, we introduce the stacked vector $\Theta^K :=((Y^K)^\top,(S^K)^\top)^\top$. By assumption, the initial condition $X(1)$, and the processes, $M(k)$ and $W(k)$, $k \in \Nat$, are mutually independent, and $X(1) \sim \mathcal{N}[\mu^X_1,\Sigma^X_1]$, $M(k) \sim \mathcal{N}[\mathbf{0},\Sigma^M]$, and $W(k) \sim \mathcal{N}[\mathbf{0},\Sigma^W]$ for some positive definite covariance matrices $\Sigma^X_1$, $\Sigma^M$, and $\Sigma^W$. Then, see \cite{Ross} for details, we have $L_1 X(1) \sim \mathcal{N}[L_1 \mu^X_1,L_1 \Sigma^X_1 L_1^{\top}]$, $L_2 M^{K-1} \sim \mathcal{N}[\mathbf{0},L_2 (I_{K-1} \otimes \Sigma^M) L_2^{\top}]$, and $L_3W^{K} \sim \mathcal{N}[\mathbf{0},L_3(I_{K} \otimes \Sigma^W)L_3^\top]$, for any deterministic matrices $L_j$, $j=1,2,3$, of appropriate dimensions. It follows that $\Theta^K =((Y^K)^\top,(S^K)^\top)^\top$ given in \eqref{stackedXY} is the sum of a deterministic vector, $\begin{bsmallmatrix} \tilde{C}_K^\top  && \tilde{D}_K^\top \end{bsmallmatrix}^\top L_K U^{K-1}$, and three independent normally distributed vectors. Therefore, $\Theta^K$ follows a multivariate normal distribution with $E[\Theta^K] = \begin{bsmallmatrix} \tilde{C}_K^\top  && \tilde{D}_K^\top \end{bsmallmatrix}^\top F_K \mu^X_1 + \begin{bsmallmatrix} \tilde{C}_K^\top  && \tilde{D}_K^\top \end{bsmallmatrix}^\top L_K U^{K-1}$. By inspection, using the expression of $\Theta^K$ in \eqref{stackedXY}, mutual independence among $X(1)$, $M(k)$, and $W(k)$, $k \in \Nat$, and the definition of $\Sigma^X_1$, $\Sigma^X_1 := E[(X(1)-\mu^X_1)(X(1)-\mu^X_1)^{\top}]$, it can be verified that the covariance matrix of $\Theta^K$, $E[(\Theta^K-E[\Theta^K])(\Theta^K-E[\Theta^K])^\top]$, is given by $\Sigma^{Y,S}_K$ in \eqref{stackeconmatrix3}. It remains to prove that the distribution of $\Theta^K$ is not degenerate, i.e., $\Sigma^{Y,S}_K>0$. Note that $\Sigma^{Y,S}_K$ in \eqref{stackeconmatrix3} can be written as
\begin{equation}\label{trans1}
\Sigma^{Y,S}_K = \begin{bmatrix} \tilde{C}_KQ\tilde{C}_K^\top + (I_{K} \otimes \Sigma^W) & \tilde{C}_KQ\tilde{D}_K^\top \\[1.5mm] \tilde{D}_KQ\tilde{C}_K^\top & \tilde{D}_KQ\tilde{D}_K^\top \end{bmatrix},
\end{equation}
with $Q := F_K\Sigma^X_1F_K^\top + T_K(I_{K-1} \otimes \Sigma^M)T_K^\top$. A necessary condition for the block matrix $\Sigma^{Y,S}_K$ in \eqref{trans1} to be positive definite is that the diagonal blocks are positive definite \cite{Horn}. The left-upper block is positive definite because $\Sigma^W>0$ (which implies $(I_{K} \otimes \Sigma^W)>0$ \cite{Horn}). The right-lower block is positive definite if $\tilde{D}_K$ is full row rank and $Q$ is positive definite. Because $D$ is full row rank by assumption, matrix $\tilde{D}_K := (I_K \otimes D)$ is also full row rank \cite[Theorem 4.2.15]{Horn2}. Note that $Q$ can be factored as follows
\[
Q  = \underbrace{\begin{bmatrix}F_K & T_K\end{bmatrix}}_{P}\underbrace{\begin{bmatrix}\Sigma^X_1 & \mathbf{0}\\ \mathbf{0} & I_{K-1} \otimes \Sigma^M\end{bmatrix}}_{Q^\prime}\begin{bmatrix}F_K & T_K\end{bmatrix}^\top.
\]
That is, $Q$ is a linear transformation of the block diagonal matrix $Q^\prime$ above. By inspection, it can be verified that matrix $P = [F_K \hspace{1mm} T_K]$, see \eqref{stackedY}, is lower triangular with identity matrices on the diagonal; thus, $P$ is invertible. It follows that $Q = PQ^\prime P^\top$ is a congruence transformation of $Q^\prime$ \cite{BEFB:94}. The later implies that $Q$ and $Q^\prime$ have the same signature \cite{BEFB:94} (equal number of positive and nonpositive eigenvalues); hence, $Q$ is positive definite if and only if the block diagonal matrices of $Q^\prime$ are positive definite. Matrices $\Sigma^X_1$ and $\Sigma^M$ are positive definite by assumption (which implies $(I_{K-1} \otimes \Sigma^M)>0$), and thus we can conclude that $Q>0$, which implies $\tilde{D}_KQ\tilde{D}_K^\top>0$ because $\tilde{D}_K$ is full row rank. Necessary and sufficient conditions for $\Sigma^{Y,S}_K>0$ are $\tilde{D}_KQ\tilde{D}_K^\top>0$ (which we have already proved) and that the Schur complement of block $\tilde{D}_KQ\tilde{D}_K^\top$ of $\Sigma^{Y,S}_K$, denoted as $\Sigma^{Y,S}_K/\tilde{D}_KQ\tilde{D}_K^\top$, is positive definite \cite[Theorem 1.12]{SchurComp}. This Schur complement is given by
\begin{align*}
&\Sigma^{Y,S}_K/\tilde{D}_KQ\tilde{D}_K^\top\\
&= (I_{K} \otimes \Sigma^W) +  \tilde{C}_K \big( Q - Q\tilde{D}_K^\top (\tilde{D}_KQ\tilde{N}_K^\top)^{-1} \tilde{N}_KQ \big) \tilde{C}_K^\top.
\end{align*}
Since matrix $(I_{K} \otimes \Sigma^W)$ is positive definite by construction, a sufficient condition for $\Sigma^{Y,S}_K/\tilde{D}_KQ\tilde{D}_K^\top>0$ is \[ Q'' := Q - Q\tilde{D}_K^\top (\tilde{D}_KQ\tilde{N}_K^\top)^{-1} \tilde{N}_KQ>0.\] Regarding $Q''$ as the Schur complement of a higher dimensional matrix $Q'''$, we can conclude that:
\begin{align*}
Q'' \geq 0 \Longleftrightarrow Q''' &:=  \begin{bmatrix} Q \\[1.5mm] \tilde{D}_KQ \end{bmatrix}Q^{-1}\begin{bmatrix} Q & Q\tilde{D}_K^\top \end{bmatrix}   \geq 0,
\end{align*}
which is trivially true because $Q^{-1}$ is positive definite since $Q>0$. Hence, $\tilde{D}_KQ\tilde{D}_K^\top$ and $\Sigma^{Y,S}_K/\tilde{D}_KQ\tilde{D}_K^\top$ are both positive definite, and thus $\Sigma^{Y,S}_K>0$. \hfill   $\blacksquare$

\bibliographystyle{IEEEtran}
\bibliography{ifacconf32}

\end{document}